# A fully Lagrangian DEM-MPS mesh-free model for ice-wave dynamics

**Rubens Augusto Amaro Junior[1,2], Andrea Mellado-Cusicahua[1], Ahmad Shakibaeinia[1,3] and Liang-Yee Cheng[2]**

[1]*Department of Civil, Geological and Mining Engineering, Polytechnique Montreal, Canada*

[2]*Department of Construction Engineering, Polytechnic School of the University of São Paulo, Brazil*

[3]*Canada Research Chair in Computational Hydrosystems, Canada*

rubens.amaro@usp.br
andrea.mellado-cusicahua@polymtl.ca
ahmad.shakibaeinia@polymtl.ca
cheng.yee@usp.br

**Abstract:** This paper develops and evaluates a novel three-dimensional fully-Lagrangian (particle-based) numerical model, based on the hybrid discrete element method (DEM) and moving particle semi-implicit (MPS) mesh-free techniques, for modeling the highly-dynamic ice-wave interactions. Both MPS and DEM belong to mesh-free Lagrangian (particle) techniques. The model considers ice-wave dynamics as a multiphase continuum-discrete system. While the MPS solves the continuum equations of free-surface flow in a Lagrangian particle-based domain, the DEM uses a multi-sphere Hertzian contact dynamic model to simulate the ice floes motion and interaction. The hybrid model predicts the motion and collision of ice floes as well as their interaction with water, boundaries, and any obstacle in their way. Considering the mesh-free Lagrangian nature of both DEM and MPS, the developed model has an inherent ability to predict the free-drift (absence of internal stress) movements of the ice floes, e.g., sliding, rolling, colliding, and piling-up, in violent free-surface flow. A small-scale and challenging experiment based on dam-break flow over dry and wet beds with floating block floes, which mimics the characteristics of an idealized jam release, has been conducted to provide useful and comprehensive quality data for the validation of the proposed model, as well as other numerical models. Experimental and numerical results of the free-surface profile and the position of the blocks are compared. The results show the ability of the model to numerically reproduce and predict the complex three-dimensional dynamic behavior of wave-ice floes interaction. Overall, this study is a first effort toward developing an ice-wave dynamics within a fully Lagrangian framework (i.e. both flow hydrodynamics and ice dynamics in the Lagrangian particle-based system), and its results can be extended to bring an in-depth understanding of the physics of the real-scale ice-wave or river ice dynamic problems in the future.

# 1    Introduction

River ice processes play an important role in cold-region hydrosystems (e.g., rivers and lakes), especially during the ice formation and breakup seasons. At the beginning of winter, when rivers become supercooled, small crystals known as frazil ice start to form and they agglomerate creating ice floes. At the end of winter, ice floes are formed due to the breakup of the solid ice covers caused by temperature and flow increase (Beltaos et al., 2000). These ice floes are carried downstream and can impact the river morphology and ecosystem and threaten riverside communities and infrastructures (e.g., bridges and hydroelectric dams, buildings, and roads) (Beltaos, 2010). Furthermore, they can jam and block the flow leading to the rise of the upstream water level and flooding. The increase in the forces acting on the jam or cover front can provoke the breakup and release of the ice jam, with a high magnitude flood wave causing the ice floes to be carried downstream while sliding, rolling, and colliding with everything at high speed damaging infrastructures and threatening human lives (Beltaos, 2010). Similar high dynamic interaction of wave and ice can also be observed in the dam-break of ice-covered reservoirs and in some sea ice processes. Predicting and understanding the processes and mechanisms involved in the dynamics of wave-ice interaction is crucial for assessing and mitigating the potential impacts. Nevertheless, due to the complexities involved in such a highly dynamic multi-physics system, these processes and mechanisms have remained largely unpredictable, especially in the case of river ice.

Due to the intrinsic complexity, simplified theoretical models of river ice dynamics or wave-ice interactions are unable to describe the complete problem. Field measurements were made to study and understand river ice (Doyle & Andres, 1979; Ford et al., 1991; Hicks, 2003; She et al., 2009; Beltaos et al., 2011; Beltaos et al., 2018) and wave-ice interaction dynamics in the contexts of ice jam processes (Jasek, 2003; Nafziger et al., 2016; Nafziger et al., 2019) and sea ice (Squire, 2007; Squire, 2018; Squire, 2020), but the data obtained were generally limited, scarce, insufficient, and expensive due to instrumental and accessibility limitations as well as the associated risks. Several past studies have used laboratory experimentation to shed light on the complex dynamics of river ice (Morse et al., 1999; Healy & Hicks, 2001; Beltaos, 2007).

More recent efforts have focused on the numerical methods, which can provide an economical alternative to simulate different scenarios that can include future changes such as regulations and/or climate change, e.g., Blackburn and She (2019). Many past numerical studies on ice dynamics (particularly for river ice) have been based on one-dimensional (1D) models. Though these models have been able to provide a general picture of the ice processes (useful for many applications), they are limited in providing a detailed undelaying physics, especially when the second and third dimension play important roles in the flow structure and ice floe motions (e.g., when dealing with highly-dynamic flows with circulations and complex geometries). In the past, several two-dimensional (2D) models have been successfully applied to simulations of general ice dynamics, including frazil ice, jamming of ice floes, ice jam release and ice cover breakup, as stated in the extensive review of Shen (2010). Subsequently, a range of three-dimensional (3D) numerical formulations using Eulerian mesh-based methods has been applied to solve ice-wave interaction problems (Wang & Meylan, 2004; Gagnon & Wang, 2012; Song et al., 2016; Bai et al., 2017; Sayeed et al., 2018). Nevertheless, dealing with the free-floating solids in free-surface flows is still a challenge due to the presence of splashing, fragmentations, merging, or multiple body interactions. Furthermore, boundary tracking or re-meshing techniques are required for mesh-based methods, increasing the computational complexity.

Due to the ability to deal with complex problems involving moving boundaries, and large deformed and fragmentation, the mesh-free particle-based (Lagrangian) methods, such as the discrete element method (DEM), introduced by Cundall and Strack (1979) for modeling material as an assembly of independent particles, the smoothed particle hydrodynamics (SPH), originally proposed to solve compressible astrophysical problems (Gingold & Monagham, 1977; Lucy, 1977) and the moving particle semi-implicit (MPS), originally developed for incompressible flows (Koshizuka & Oka, 1996), opened new perspectives for modeling fluid-solid interaction problems in recent years.

In the context of ice dynamics, recent efforts on accurate simulation have resulted in two classes of numerical techniques, based on either continuum description or discrete description of ice floes. While the accuracy of the continuum description largely depends on the accuracy of the ice rheological models, the discrete models have shown to be able to describe the underlying physics in detail. Among discrete techniques, the particle-based method DEM have been extensively used to study ice mechanics. The studies of Hopkins and Hibler (1991) and Løset (1994; 1994), in which ice floe fields were considered as systems of 2D discs, are pioneering works on the application of DEM to describe the ice behaviour. Hopkins (2004) used a 2D polygonal DEM approach, based on previous DEM work (Daly & Hopkins, 2001; Hopkins & Tuthill, 2002), to simulate the sea ice floes displacement caused by wind stress and water drag. Ice floes thickness were modeled using an algorithm proposed by Ebert and Curry (1993), in which the ice dynamics incorporates a lead parameterization that takes into account a minimum lead fraction, the absorption of solar radiation in and below the leads, the lateral accretion and ablation of the sea ice and a prescribed sea ice divergence rate. Hopkins and Daly (2003) developed a similar DEM model that includes not only the drag force but also the buoyancy and pressure force. This model combines a 3D DEM with a 1D hydraulic model whose transverse and vertical geometry and flow data are averaged. In Hopkins and Shen (2001) and Dai et al. (2004), a 3D DEM model was used to study the dynamics of pancake-ice, i.e., the circular floes formed during ice growth in a wave field. They demonstrated that the final ice cover thickness due to the rafting process is a function of wave amplitude, wavelength and floe diameter. Another model developed by Stockstill et al. (2009) took a step forward and combined the 3D DEM with depth-averaged 2D shallow-water equations solved using a finite element (FE) scheme. This model successfully simulated ice transportation and accumulation, as well as the interaction between ice and hydraulic structures. In Karulin and Karulina (2011), the moored tanker's behaviour in broken ice was investigated by the DEM approach. The ice rubbling process was simulated by the finite-discrete element method in Paavilainen and Tuhkuri (2013), where an ice sheet and blocks were modeled using a finite element method (FEM), and the ice rubble pile was modeled by a discrete element scheme. The pressure distributions during select peak load events were numerically investigated, with special attention on the effects of a loose rubble pile in front of the structure on the pressure distribution. Ji et al. (2015) applied the DEM to analyse the influences of ice velocity, ice thickness, and conical angle on ice loads in conical offshore structures. A dilated polyhedral DEM was developed by Liu and Ji (2018), and numerical simulations of the interaction between ice floes and the floating structure Kulluk were validated against field data. Herman et al. (2019) conducted numerical studies on wave attenuation through ice-floe fields, and used DEM to account for the energy dissipation due to ice floe collisions. In Gong et al. (2019), the resistance of a ship in ridges of equal depth but different widths has been studied using a 3D DEM. A good review of different types of DEM approaches and their applications on ice-related problems can be found in Metrikin and Løset (2013), Tuhkuri and Polojärvi (2018), and Xue et al. (2020). Most of the numerical flow models adopted in these works have been based on simplified depth-averaged hydrodynamics, or potential flow theory. Hence, they are not able to reproduce the highly dynamic ice-wave interactions.

Regarding the numerical simulation of fluid-solid interaction, it can be modeled by a multiphase continuum-discrete system in particle-based methods, in which the fluid phase is based on SPH or MPS, where continuum conservation equations are solved over a set of particles, containing the continuum field variables, that move in the Lagrangian system (Shakibaeinia & Jin, 2010), while the motion of the solid phase is solved using the DEM. Among the various representations adopted in the DEM to model solids of different scales or complex shapes, two representations are used in several works due to their efficiency and robustness:

1. Granular materials modeled as individual discrete particles, i.e., both solid and fluid phases have a similar scale;

2. Complex-shaped or large-scale bodies modeled as a collection of particles, the so-called multi-sphere method (Favier et al., 1999).

Based on the former representation, the DEM-SPH/MPS approach has been applied to engineering problems such as the interaction between fluid and solid particles in a cylindrical tank (Sakai et al., 2012), particle sedimentation (Robinson et al., 2014), ice dynamics (Robb et al., 2016), sediment transport in the swash zones (Harada et al., 2019), non-Newtonian solid-liquid interaction with the presence of free-surface flow (Li et al., 2019), among others. Concerning the latter representation, some works proposed the numerical approach combining DEM and SPH/MPS mainly to simulate rigid bodies transport under free-surface flow (Canelas et al., 2016), (Guo et al., 2017), (Amaro Jr et al., 2019), (Ji et al., 2019), (Wang et al., 2019). Moreover, to the author's knowledge, only Kawano and Ohashi (2015) and Zhang et al. (2019) conducted numerical simulations of ice-wave interaction using particle-based methods. Kawano and Ohashi (2015) reproduced the process of growth and accumulation of crystal nuclei and the formation process of the layer of fine crystals adopting a particle agglomeration with MPS. The sea ice crystal growth from the sea surface was simulated by the Voronoi dynamics technique (Ohashi et al., 2004). However, only the calm sea condition was considered, and a collision model was not presented. In Zhang et al. (2019), experimental measurements related to the kinematic and flexural responses of a deformable ice floe induced by water waves were compared to results obtained from numerical simulations using the SPH. They proposed a fluid-solid interaction scheme based on the momentum equations discretized by the SPH approximations. Nonetheless, only 2D cases with one ice floe were simulated, and they did not introduce any solid collision model.

In this paper, a novel 3D fully Lagrangian model able to handle ice-wave interaction is proposed, which couples MPS and DEM, for continuum (water) and discrete (ice) phases, respectively. Although DEM has been extensively used in the past for the ice dynamic simulations, this is the first effort toward an efficient fully Lagrangian model that combines DEM with a continuum particle method (e.g., MPS) to make it capable of simulating the highly dynamic 3D ice-wave systems. In the MPS method, the differential operators of the continuum mechanics are replaced by weighted average discrete operators on irregular nodes, and a semi-implicit algorithm is applied to solve the governing equations. Furthermore, MPS is very effective for the simulations of incompressible flow involving large deformation of free surfaces, fragmentation, and merging, or involving complex shaped bodies, large deformation or motion of boundaries, multi-bodies, multi-phase flows, and multi-physic problems. The DEM is coupled with the MPS method and is used to simulate the solid ice phase, here simplified by rigid body. Here the DEM is adopted following the original idea of Koshizuka et al. (1998), the so-called passively moving solid (PMS) model, in which the shell of a rigid body is represented by a cluster of particles whose relative positions remain unchanged, similar to the multi-sphere technique. In order to address numerical instabilities due to non-smooth solid walls modeled by particles, an approach based on the faces of the bodies and contact force computed using the normal vectors of solid walls (Amaro Jr et al., 2019) is adopted here. This study uses a high resolution of particles for both DEM and MPS, e.g., a single ice flow is represented with more than 5000 particles, leading to computationally expensive simulations. Therefore, this study implements the model on a shared-memory parallel code (to use the power of multicore processors) and limits the test cases to simple geometries with a few ice floes simplified by rigid bodies, i.e., internal ice stress and broken ice behaviour are neglected, and therefore the dynamics are close to a state of free-drift motions governed by the flow. The physical and mathematical aspects of free-drift motions of ice are well discussed in Leppäranta (2011). Future extension of the model for the real-scale problems (with hundreds of floes) can be done using a massively parallel code (see Fernandes et al. (2015)), and by incorporating multi-resolution techniques (Chen et al., 2016; Tang et al., 2016; Shibata et al., 2017; Tanaka et al., 2018; Khayyer et al., 2019; Sun et al., 2019), in which high-resolution is used only near the local critical areas, while low-resolution is used in the far-field.

Moreover, considering the currently lacking comprehensive data for the detailed validation of ice-wave interaction dynamics, a small-scale yet challenging experiment based on dam-break flow over dry and wet beds with floating block floes (ice dynamics simplified by dynamics of rigid bodies), which somehow mimics the characteristics of an idealized jam release, has been conducted to provide quality and quantitative data for the validation of the proposed model and investigation of the highly nonlinear phenomenon. Though this experiment is a simplification of a real jam release, it provides the basic solid-wave interactions needed to validate the proposed DEM-MPS simulation. The imagery data from two high-speed cameras have been analyzed and used to track the highly dynamic motion of the blocks and water surface profiles.

## 2   Experimental setup

The experiments consist of a series of dam-break case scenarios with floating blocks that offer similar characteristics to jam releases and breaches of ice-covered reservoirs in terms of having high magnitude waves with floating ice parcels. Reported experiments in the literature of wave-ice interaction generally cover two scenarios:

i.) one or two ice floes interacting with waves (McGovern & Bai, 2014; Yiew et al., 2016; Yiew et al., 2017);

ii.) hundreds of ice floes subjected to waves and colliding with each other (Dai et al., 2004; Bennetts & Williams, 2015) and also with towing carriage (Hu & Zhou, 2015; Luo et al., 2018).

In the present work, our goal is to provide an experimental benchmark test to fill the gap between these two situations. In this way, the present test cases were selected based on their high-dynamic nature, simple geometry, reduced number of ice floes, and valuable data that they can provide for the numerical model validation.

### 2.1   Experimental apparatus

The experiments have been conducted at the hydraulic laboratory of the École Polytechnique of Montreal. A prismatic plexiglass tank having interior dimensions of 70 cm length, 15 cm width, and 30 cm height, was used here, see Fig. 1(a). A removable gate of 0.5 cm thickness divides the tank into two parts. The upstream part has a length of $L_{up}$ = 15 cm and water height of $H_{up}$ = 15 cm, and the downstream part has a length of $L_{do}$ = 55 cm and variable water heights $H_{do}$, respectively, see Fig. 1(b). Blocks of artificial ice material float on the surface of the upstream reservoir. For analysis purposes, a reference frame, having the origin at point A, has been adopted. To replicate the dam-break event, the gate was removed (lifted) almost instantaneously through a pulley-weight mechanism, as shown in Fig. 1(b). The sides of the gate were sealed with flat and flexible vinyl strips and high-vacuum grease to ensure the reservoir's water tightness. Two high-speed cameras were used to track the longitudinal and vertical motions. A high-speed camera (FASTCAM Mini WX100), capable of taking 1080 frames per second (fps) with a 2048 × 2048 resolution, was placed on top of the tank, and another one (Sony DSC-RX100M5A) with 480 fps and 1292 × 436 resolution was placed at the front of the tank. The light source was a high-frequency AOS Offboard 150W Led, which eliminates flicker issues.

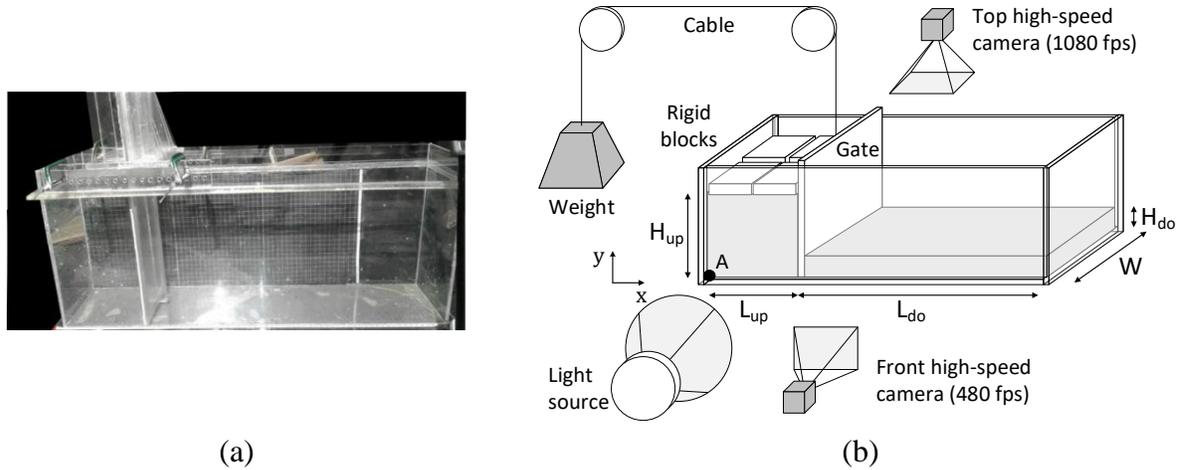

(a)　　　　　　　　　　　　　　　　(b)

Fig. 1. (a) Prismatic plexiglass tank used in the dam-break experiments. (b) Schematic drawing of the tank, water levels, gate, rigid blocks, pulley-weight mechanism, cameras disposition, and light source.

## 2.2　Experimental material

Blocks of white polypropylene, with a density of approximately 868 kg/m³, were used as the artificial ice floes. However, it shall be mentioned that a higher density around 0.92 kg/m³, can provide a better estimation for river ice floes, i.e., salinity ~0 ppt (parts per thousand), see Fig. 3 in Timco and Frederking (1996). The friction coefficient required for the numerical simulations was determined experimentally. The static friction coefficient between blocks was measured using the setup shown in Fig. 2. Two sets of polypropylene strips were attached, where the top one was free to move, and the bottom one was fixed to the table. Knowing that the blocks are wet during the experiment, the strips were soaked wet before applying a horizontal force. By putting additional weight above the free set of strips, the normal force was defined, and the shear force was obtained by measuring the force required to move the block horizontally. The different masses put on top of the blocks, the average measured shear force, and the correspondent friction coefficients calculated as the ratio of the shear force to the normal are shown in Appendix A. The median static friction coefficient found for the wet blocks and used in the numerical simulations was $\mu_s = 0.412 \pm 0.050$. It should be mentioned that the present test did not consider the effect of sliding speed, temperature, and surface tension on the friction coefficient. Furthermore, the static friction coefficient might not be the best parameter to use in dynamic simulations, but the relevance of the friction forces seems to be small compared to the other solid collision forces in the present study. Nonetheless, to the authors' knowledge, a wide range of values [0.1, 0.6], given by a function of the velocity of the interaction, has been used for the actual ice-ice friction coefficient (Timco & Weeks, 2010; van den Berg et al., 2019). In this way, we intended to provide a value of reference for the friction coefficient for the material used here, namely polypropylene, with the available laboratory instrumentation in the hydraulic laboratory of the École Polytechnique Montreal.

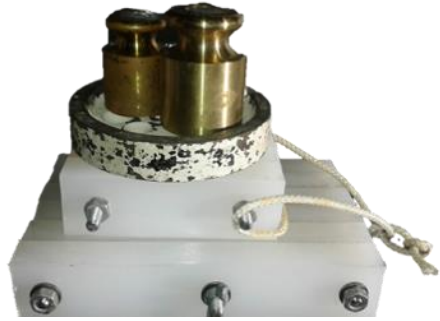

Fig. 2. Sets of polypropylene strips and weights.

## 2.3 Experimental case scenarios

Eight case scenarios, which differ in the number of polypropylene blocks and the downstream water heights, were carried out as summarized in Table 1. In the first four cases, 4 equal-size polypropylene blocks were used. Each block has dimensions (width x length x thickness) 7.25 x 7.25 x 1.95 ± 0.06 cm and a mass of 0.089 ± 0.001 kg. The other four cases used 9 polypropylene blocks of 4.80 x 4.80 x 1.95 ± 0.06 cm, each having a mass of 0.039 ± 0.001 kg. Four downstream water heights, i.e., $H_{do}$ = 0, 1.0, 2.5 and 5.0 cm, were tested. Each case was repeated at least three times to check the reliability, and the three repeated experiments of each case are denominated EXP1, EXP2, and EXP3.

Table 1. Experimental case conditions. Number of blocks, upstream ($H_{up}$) and downstream ($H_{do}$) water height.

| Case | Number of blocks | $H_{up}$ (cm) | $H_{do}$ (cm) |
|------|------------------|---------------|---------------|
| 1    |                  |               | 0.0           |
| 2    | 4                |               | 1.0           |
| 3    |                  |               | 2.5           |
| 4    |                  | 15            | 5.0           |
| 5    |                  |               | 0.0           |
| 6    | 9                |               | 1.0           |
| 7    |                  |               | 2.5           |
| 8    |                  |               | 5.0           |

## 2.4 Data acquisition

The evolution of the water surface profile and displacement of ice blocks were determined by processing high-speed imagery data from the top and front cameras. The displacements of the blocks in x and y directions were determined using the high-speed videos obtained from the top and front cameras, respectively. For this purpose, a free video analysis software called Tracker (Brown, 2019) was used to automatically track the center of mass (CM) of each block, as shown in Fig. 3, using a template matching algorithm. The water surface profile was obtained by digitizing images extracted from the videos taken by the front camera.

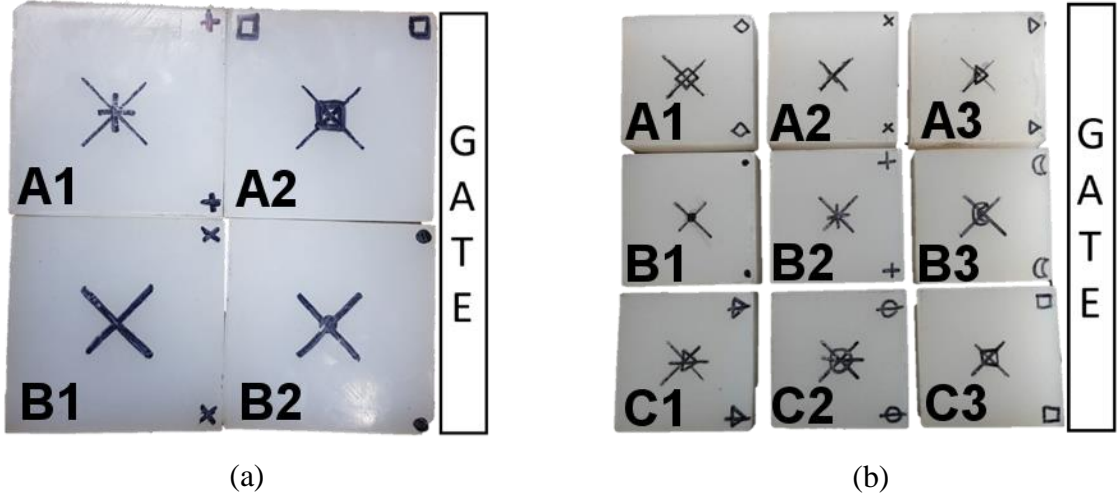

<div align="center">(a)                        (b)</div>

Fig. 3. Initial arrangement of the cases with (a) 4 blocks and (b) 9 blocks.

## 2.5   Gate motion

The time series of the vertical gate motion for the 24 experiments are presented in Fig. 4. According to Lauber and Hager (1998), the gate motion satisfies the criterion of a sudden removal if the removal period $t_r$ (time required for a fluid particle located at the top of the fluid column to reach the bottom of the tank) is smaller than $\sqrt{2H_{up}/g}$. From the video records, the gate removal duration, i.e., the instant when the gate's lower edge reaches the dam filling height, was around 0.40 s and did not satisfy the criterion $t_r < \sqrt{2 \times 0.15/9.81} \rightarrow t_r < 0.175$ s. Therefore, the present experiments can not be considered as sudden dam breaks. Furthermore, since the gate motion has a significant influence on the water collapse process and dam-break results, it cannot be neglected in the numerical model. Nevertheless, we performed some simulations using: i) the behavior of the vertical gate motion observed in the experiments (see Fig. 4), i.e., a variable vertical velocity; and ii) a constant vertical velocity so that the trajectory of the gate becomes linear. Negligible differences in the wave-profile evolutions as well as the solid motions were observed between the experimentally measured and the numerically computed results when using the varying or constant gate velocity. Thus, the constant gate velocity, which can be easily assigned as a boundary condition in a wider range of numerical solvers, was chosen based on the median of the average gate velocity of the 24 cases ($v_g = 0.4$ m/s) and adopted in the numerical simulations. The median value of gate velocity and the relevant confidence intervals are shown in Fig. 4.

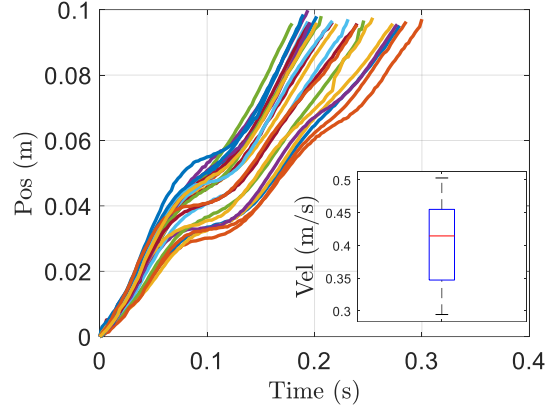

Fig. 4. Time series of the vertical gate motion for all 24 experiments and the median value of gate velocity.

## 3 Governing equations

The physical system includes the interaction of a continuum phase (water) and a discrete phase (solid ice blocks). The governing equations describing the dynamics of the continuum phase (flow of an incompressible viscous fluid) are expressed by the conservation laws of mass and momentum, which in their Lagrangian form are as follows:

$$\frac{D\rho_f}{Dt} + \rho_f \nabla \cdot \mathbf{u}_f = 0 \, , \tag{1}$$

$$\frac{D\mathbf{u}_f}{Dt} = -\frac{\nabla P}{\rho_f} + \nu_f \nabla^2 \mathbf{u}_f + \boldsymbol{\mathcal{F}}_b \, , \tag{2}$$

where $\rho_f$ is the fluid density, $\mathbf{u}_f$ stands for the fluid velocity vector, $P$ denotes the pressure, $\nu_f$ represents the kinematic viscosity and $\boldsymbol{\mathcal{F}}_b$ is the external body force per unit mass vector, namely the gravitational acceleration vector $\mathbf{g}$ in the present work.

For the discrete phase (solid blocks), the governing equations of motion are those of translational and rotational motion given by:

$$m_s \frac{D\mathbf{u}_s}{Dt} = \mathbf{F}_h + \mathbf{F}_g + \mathbf{F}_c = -\iint_{\mathcal{A}} P d\mathbf{a} + m\mathbf{g} + \mathbf{F}_c \, , \tag{3}$$

$$\mathbf{I} \cdot \frac{D\boldsymbol{\omega}_s}{Dt} + \boldsymbol{\omega}_s \times (\mathbf{I} \cdot \boldsymbol{\omega}_s) = \mathbf{T}_h + \mathbf{T}_c = -\iint_{\mathcal{A}} \mathbf{r}_s \times P d\mathbf{a} + \mathbf{T}_c \, , \tag{4}$$

where $m_s$ is the total mass of the rigid body, $\mathbf{u}_s$ represents the velocity vector at the CM of the rigid body, $\mathbf{I}$ is the inertia matrix and $\boldsymbol{\omega}_s$ stands for the angular velocity vector about the principal axes of the rigid body. The hydrodynamic forces on the rigid surface $\mathbf{F}_h$, gravitational force $\mathbf{F}_g$, contact forces between the rigid bodies $\mathbf{F}_c$, hydrodynamic torque $\mathbf{T}_h$ and contact torque $\mathbf{T}_c$ are taken into consideration for the motion of rigid bodies. The vector $\mathbf{r}_s$ denotes the position vector from the CM of the rigid body and $d\mathbf{a}$ is the face vector of the rigid body surface, whose magnitude is the area of the discrete face and direction is normal to the body surface. Focusing the impulsive hydrodynamic loads on the rigid solid, the contribution of fluid shear forces was assumed to be negligible in the present study. It is important to highlight that it is a practical limitation of the present model since shear forces would be very important in ice jam formation and release events.

## 4    Numerical methods

Here, the moving particle semi-implicit (MPS) method and discrete element method (DEM) are used to solve the governing equation of the continuum-phase (water) and discrete phase (solid ice blocks), respectively.

### 4.1    *Moving Particle Semi-implicit (MPS)*

As a particle method, MPS represents the continuum with a set of mobile particles (without any connectivity) over which the flow governing equations are solved. Here, a semi-implicit algorithm, which divides each time step into prediction and correction steps, is used for the temporal integration of the governing equations. At first, predictions of the velocity and position of a fluid particle $i$ are carried out explicitly by using viscosity and external forces terms of the momentum conservation:

$$\mathbf{u}_i^* = \mathbf{u}_i^t + \Delta t \big[ \nu_f \langle \nabla^2 \mathbf{u} \rangle_i + \boldsymbol{\mathcal{F}}_b \big]^t, \tag{5}$$

$$\mathbf{r}_i^* = \mathbf{r}_i^t + \Delta t \mathbf{u}_i^*. \tag{6}$$

Here, the superscript * refers to the prediction step. The approximation of spatial derivations in MPS is based on the kernel smoothing (weighted averaging) process. The MPS approximation of Laplacian of velocity over neighboring particles $\Omega_i$ is given by:

$$\langle \nabla^2 \mathbf{u} \rangle_i = \frac{2d}{\lambda_i pnd^0} \sum_{j \in \Omega_i} (\mathbf{u}_j - \mathbf{u}_i) \omega_{ij}, \tag{7}$$

where $d = 1, 2$ or $3$ is the number of spatial dimensions, $pnd^0$ stands for the particle number density of a fully compact support with an initial cubic arrangement of particles, $\omega_{ij}$ represents a weight (kernel) function and $\lambda_i$ is a correction parameter by which the variance increase is adjusted to be equal to the analytical solution. Here, the widely used rational weight function (Koshizuka & Oka, 1996) is used:

$$\omega_{ij} = \begin{cases} \dfrac{r_e}{\|\mathbf{r}_{ij}\|} - 1 & \|\mathbf{r}_{ij}\| \leq r_e \\ 0 & \|\mathbf{r}_{ij}\| > r_e \end{cases}, \tag{8}$$

where $r_e$ is the effective radius that limits the range of influence and $\|\mathbf{r}_{ij}\| = \|\mathbf{r}_j - \mathbf{r}_i\|$ is the distance between the particles $i$ and $j$. As demonstrated by Koshizuka and Oka (1996), more accurate and stable computations can be achieved with $r_e \in [1.8l^0, 3.1l^0]$ in 3D problems, where $l^0$ represents the initial distance between two adjacent particles. Therefore, in the present work, $r_e = 2.1l_0$ is adopted for the calculations of particle number density and all differential operators. For an extensive and quantitative study of the influence of $r_e$ on the numerical accuracy, the reader is referred to the work of Duan et al. (2019). The summation of the weight of all the particles in the neighborhood of particle $i$ is defined as its particle number density

$$pnd_i = \sum_{j \in \Omega_i} \omega_{ij}, \tag{9}$$

which is proportional to the fluid density.

The correction parameter $\lambda_i$ is defined as:

$$\lambda_i = \frac{\sum_{j \in \Omega_i} \|\mathbf{r}_{ij}\|^2 \omega_{ij}}{\sum_{j \in \Omega_i} \omega_{ij}}, \tag{10}$$

After the prediction of velocities and positions of the fluid particles, a collision model is applied to fluid particles located at the free surface or some inner particles with few neighbors to the proper calculation of the discrete differential operators, avoiding clustering of particles, and the contribution $\Delta\mathbf{u}^*$ is added:

$$\mathbf{u}_i^{**} = \mathbf{u}_i^* + \Delta\mathbf{u}_i^*, \tag{11}$$

$$\mathbf{r}_i^{**} = \mathbf{r}_i^* + \Delta t \Delta\mathbf{u}_i^*, \tag{12}$$

where $\Delta\mathbf{u}_i^*$ can be calculated from:

$$\Delta\mathbf{u}_i^* = \begin{cases} \sum_{j \in \Omega_i} \dfrac{(1+\alpha_2)}{\alpha_3} \dfrac{\mathbf{r}_{ij}^* \cdot \mathbf{u}_{ij}^*}{\|\mathbf{r}_{ij}^*\|} \dfrac{\mathbf{r}_{ij}^*}{\|\mathbf{r}_{ij}^*\|} & \|\mathbf{r}_{ij}^*\| \leq \alpha_1 l^0 \quad \text{and} \quad \mathbf{r}_{ij}^* \cdot \mathbf{u}_{ij}^* < 0 \\ 0 & \text{otherwise} \end{cases}. \tag{13}$$

If the neighbor $j$ is a fluid particle, then the coefficient $\alpha_3 = 2$ otherwise $\alpha_3 = 1$. According to Lee et al. (2011), values of the coefficient that define the collision distance $\alpha_1 \geq 0.8$ and coefficient of restitution $\alpha_2 \leq 0.2$ increase the spatial stability in simulations.

Then the pressures of fluid and wall particles are obtained implicitly by solving a linear system of pressure Poisson equation (PPE) (Koshizuka et al., 1999; Ikeda et al., 2001):

$$\langle \nabla^2 P \rangle_i^{t+\Delta t} - \frac{\rho_f}{\Delta t^2} \alpha_c P_i^{t+\Delta t} = \gamma \frac{\rho_f}{\Delta t^2} \left( \frac{pnd^0 - pnd_i^{**}}{pnd^0} \right), \tag{14}$$

where $\Delta t$ is the time step, $pnd_i^{**}$ is the particle number density calculated after the prediction and collision processes, $\alpha_c$ is the coefficient of artificial compressibility, and $\gamma$ is the relaxation coefficient. Both $\alpha_c$ and $\gamma$ are used to improve the stability of the computation method. Eq. (14) represents a linear system, characterized by a sparse matrix, in which a higher coefficient $\alpha_c$ makes the diagonal elements of the matrix bigger, rending it very useful for computational stabilization. The relaxation coefficient $\gamma$ is adopted to enforce the incompressibility condition in a robust way while mitigating high-frequency pressure oscillations in the discrete PPE. Nevertheless, both $\alpha_c$ and $\gamma$ should be chosen appropriately in order to avoid non-physical fluid behavior. Typically, the ranges $\alpha_c \in [10^{-9}, 10^{-8}] \, ms^2/kg$ and $\gamma \in [0.001, 0.05]$ provide stable simulations (Arai et al., 2013; Shibata et al., 2015; Duan et al., 2019; Tsukamoto et al., 2020). The experience of the authors has shown that the coefficients $\alpha_c = 10^{-8} \, ms^2/kg$ and $\gamma = 0.01$ give satisfactory results, and therefore they were used for all simulations herein.

The Laplacian of pressure is approximated by:

$$\langle \nabla^2 P \rangle_i = \frac{2d}{\lambda_i pnd^0} \sum_{j \in \Omega_i} (P_j - P_i) \omega_{ij}. \tag{15}$$

Finally, the velocity of the fluid particles ($\mathbf{u}_i^{t+\Delta t}$) is updated by using the pressure gradient term of the momentum conservation (see Eq. (2)) and the updated positions $\mathbf{r}_i^{t+\Delta t}$ are obtained:

$$\mathbf{u}_i^{t+\Delta t} = \mathbf{u}_i^{**} - \frac{\Delta t}{\rho_f} \langle \nabla P \rangle_i^{t+\Delta t}, \tag{16}$$

$$\mathbf{r}_i^{t+\Delta t} = \mathbf{r}_i^{**} + \Delta t \left( \mathbf{u}_i^{t+\Delta t} - \mathbf{u}_i^{**} \right). \tag{17}$$

To prevent instability issue induced by attractive pressure and to reduce the effect of nonuniform particle distribution, the first-order pressure gradient was adopted here as (Wang et al., 2017):

$$\langle \nabla P \rangle_i = \left[ \sum_{j \in \Omega_i} \omega_{ij} \frac{\mathbf{r}_{ij}}{\|\mathbf{r}_{ij}\|} \otimes \frac{\mathbf{r}_{ij}^T}{\|\mathbf{r}_{ij}\|} \right]^{-1} \left[ \sum_{j \in \Omega_i} \frac{P_j - \hat{P}_i}{\|\mathbf{r}_{ij}\|^2} \mathbf{r}_{ij} \omega_{ij} \right], \tag{18}$$

where $\hat{P}_i$ is the minimum pressure between the neighborhood of the particle $i$, i.e., $\hat{P}_i = \min_{j \in \Omega_i}(P_j, P_i)$.

## 4.2    Discrete Element Method (DEM)

Here, an approach similar to the multi-sphere DEM technique is used in which the shells of a solid body (here rigid cuboids) are represented by a collection of wall and internal spherical particles whose relative positions remain unchanged, following Koshizuka et al. (1998), see Fig. 5.

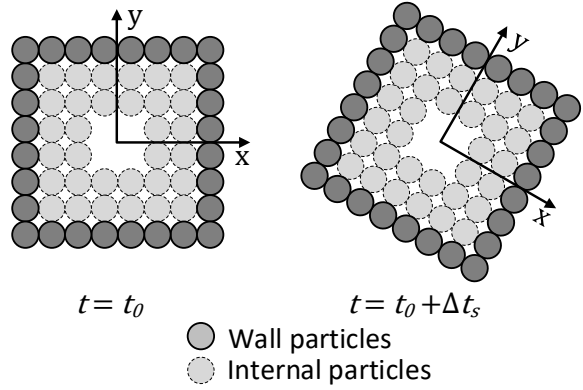

Fig. 5. Rigid body represented by a collection of wall and internal particles in 2D space.

The overall contact force on the rigid body is calculated by adding the contributions due to the normal $\mathbf{F}_n$ and tangential $\mathbf{F}_t$ contact forces between wall particles belonging to different bodies. Here, subscript $n$ stands for the normal component, and $t$ denotes the tangential component.

The normal forces between the pair of the closest particles $k$ and $l$ belonging to different bodies can be described following a non-linear Hertz's elastic contact theory (Johnson, 1985):

$$\mathbf{F}_{n,kl} = k_{n,kl}\delta_{kl}^{3/2}\mathbf{n}_c + c_{n,kl}\delta_{kl}^{1/4}\dot{\delta}_{kl}\mathbf{n}_c \,, \tag{19}$$

where $k_{n,kl} = 4/3E_{kl}\sqrt{l_{kl}^0}$ is the normal stiffness constant of pair $kl$, $\delta_{kl}$ is the overlap (penetration) between two wall particles belonging to two different bodies, $\dot{\delta}_{kl}$ is the rate of normal penetration, $c_{n,kl} = \xi_n\sqrt{6m_{kl}E_{kl}\sqrt{l_{kl}^0}}$ is the normal damping constant and $\mathbf{n}_c$ is the contact normal vector defined as follows:

1. If particle $l$ is *face*, the normal of $l$ is used as $\mathbf{n}_c$.

2. If particle $l$ is not face, but $k$ is *face*, the reverse of the normal of $k$ is used as $\mathbf{n}_c$.

3. None of the particles are face, the reverse of the distance vector $\mathbf{r}_{kl}/|\mathbf{r}_{kl}|$ is used as $\mathbf{n}_c$.

As the geometry of a cuboid is enough to model all rigid bodies considered herein, the number of neighbor solid particles is used as the criterion to identify if a particle belongs to a face.

The damping ratio $\xi_n$ must be high enough to obtain a non-oscillatory motion of the rigid solids while not too high so that too much energy is dissipated unnaturally during the collision, whereas $E_{kl}$, $m_{kl}$ and $l_{0,kl}$ are obtained as:

$$E_{kl} = \frac{E_k E_l}{(1-v_l^2)E_k + (1-v_k^2)E_l}, \quad m_{kl} = \frac{m_k m_l}{m_k + m_l}, \quad l_{kl}^0 = \frac{l_k^0 l_l^0}{l_k^0 + l_l^0}, \tag{20}$$

where $E_k$, $E_l$, $v_k$ and $v_l$ are the Young's modulus and the Poisson's ratio of particles $k$ and $l$, respectively. In the present work, a single spatial resolution is used for all domain, i.e., the initial distance between two adjacent particles $l^0$ is the same for the particles $k$ and $l$, thus leading to $l_{kl}^0 = l^0/2$. Moreover, if the $l$-th particle belongs to a fixed rigid wall, then $m_l \to \infty$, implying that $m_{kl} = m_k$.

Tangential forces can be obtained from the Coulomb friction law given by Eq. (21) and a sigmoidal function is used to make the forces continuous in the origin regarding the tangential relative velocity $\dot{\delta}_{kl}^t$ (Vetsch, 2011). The tangential forces can also be obtained using a linear model of repulsive and damped forces given by Eq. (22).

$$\mathbf{F}_{t,kl}^C = \mu_{kl}\|\mathbf{F}_n\|\tanh(8\dot{\delta}_{kl}^t)\,\mathbf{t}_c\,, \tag{21}$$

$$\mathbf{F}_{t,kl}^L = k_{t,kl}\delta_{kl}^t\mathbf{t}_c + c_{t,kl}\dot{\delta}_{kl}^t\mathbf{t}_c\,. \tag{22}$$

Here, $\mu_{kl}$ is the friction coefficient for the pair of particles $k$ and $l$, $\delta_{kl}^t$ is the relative sliding, $\dot{\delta}_{kl}^t$ is the tangential relative velocity, $\mathbf{t}_c$ is the tangential contact vector, and $k_{t,kl} = 2/7k_{n,kl}\sqrt{l^0}$ and $c_{t,kl} = 2/7c_{n,kl}(l^0)^{1/4}$ (Hoomans, 2000) are the spring-damping coefficients. The superscripts $C$ and $L$ correspond to Coulomb and linear forces, respectively.

In the present work, when the neighbor particle $l$ belongs to a forced (prescribed motion) or fixed solid, the friction coefficient $\mu_l$ of the neighbor particle $l$ is adopted for $\mu_{kl}$, otherwise $\mu_{kl}$ is obtained as the mean value:

$$\mu_{kl} = \left(\frac{\mu_k^\theta + \mu_l^\theta}{2}\right)^\theta, \tag{23}$$

similar to the interaction viscosity between two particles given in Shakibaeinia and Jin (2012). In this work, we used the arithmetic mean for all simulations, i.e., $\theta = 1$.

The force with smaller absolute value is used as tangential force during the simulation, i.e.:

$$\mathbf{F}_{t,kl} = \begin{cases} \mathbf{F}_{t,kl}^C & \|\mathbf{F}_{t,kl}^C\| \leq \|\mathbf{F}_{t,kl}^L\| \\ \mathbf{F}_{t,kl}^L & \|\mathbf{F}_{t,kl}^C\| > \|\mathbf{F}_{t,kl}^L\| \end{cases}. \tag{24}$$

After calculating all contacts between the pairs of the closest particles belonging to different bodies, e.g., bodies $p$ and $q$, the resultant contact forces for a rigid body $p$ can be calculated by:

$$\mathbf{F}_{n,p} = \sum_{q \in NB} \left( \sum_{m \in NP} (\mathbf{F}_{n,kl})_m \right), \quad \mathbf{F}_{t,p} = \sum_{q \in NB} \left( \sum_{m \in NP} (\mathbf{F}_{t,kl})_m \right), \tag{25}$$

where $NP$ is the number of pairs of the closest particles ($m$) belonging to the shell of the rigid bodies $q \in [1, NB]$, and $NB$ is the number of bodies in contact with body $p$.

The torque that acts on the CM of the solid generated by the contact forces is evaluated as:

$$\mathbf{T}_{c,kl} = (\bar{\mathbf{r}}_{kl} - \mathbf{r}_{CM}) \times (\mathbf{F}_{n,kl} + \mathbf{F}_{t,kl}), \tag{26}$$

where $\bar{\mathbf{r}}_{kl} = (\mathbf{r}_k + \mathbf{r}_l)/2$ represents the average position vector of particle $k$ and $l$ and $\mathbf{r}_{CM}$ symbolizes the position vector of the CM. The average position $\bar{\mathbf{r}}_{kl}$ is used to ensure the conservation of angular moment. Accordingly, after calculating all torques due to contacts between particles, the contact torques for each rigid body $p$ can be obtained by:

$$\mathbf{T}_{c,p} = \sum_{q \in NB} \left( \sum_{m \in NP} (\mathbf{T}_{c,kl})_m \right). \tag{27}$$

A detailed description of the solid-solid collision model can be found in the previous work (Amaro Jr et al., 2019).

### 4.3    Boundary conditions

The rigid boundaries are modeled using three layers of particles. The particles that form the layer in contact with the fluid are denominated wall particles, of which the pressure is computed by solving PPE together with the fluid particles. The particles that form two other layers are called dummy particles in MPS (fixed walls and forced solids) and inner particles in DEM (free solids), which are used to assure the correct particle number density calculation of the wall particles. The pressure is not calculated in the dummy or internal particles. As a boundary condition of rigid walls, the null relative velocity between the fluid and the wall is imposed. The Dirichlet pressure boundary condition is imposed on the particles identified as free-surface ones and is considered during the implicit step of the method. In order to identify the free-surface particles, the neighborhood particles centroid deviation (NPCD) method (Tsukamoto et al., 2016) is adopted here. In the NCPD technique, a particle is defined as free-surface one, and its pressure is set to zero if:

$$\begin{cases} n_i < \beta \cdot n^0 \\ \sigma_i > \varrho \cdot l^0 \end{cases}. \tag{28}$$

The deviation $\sigma_i$ is written as:

$$\sigma_i = \frac{\sqrt{\left(\sum_{j\in\Omega_i}\omega_{ij}x_{ij}\right)^2 + \left(\sum_{j\in\Omega_i}\omega_{ij}y_{ij}\right)^2 + \left(\sum_{j\in\Omega_i}\omega_{ij}z_{ij}\right)^2}}{\sum_{j\in\Omega_i}\omega_{ij}}, \tag{29}$$

where $x_{ij} = (x_j - x_i)$, $y_{ij} = (y_j - y_i)$ and $z_{ij} = (z_j - z_i)$.

According to Koshizuka and Oka (1996), the constant $\beta$ should be chosen between 0.80 and 0.99 and Tsukamoto et al. (2016) suggests a $\varrho$ higher than 0.2. The NPCD method improves the stability and accuracy of the pressure computation by eliminating spurious oscillations due to misdetection of free-surface particles.

### 4.4   DEM and MPS coupling

The resultant hydrodynamic force ($\mathbf{F}_h$) and torque ($\mathbf{T}_h$), see Eq. (3), acting on the rigid bodies are calculated by integrating the pressures ($P_k$), computed by solving the PPE (Eq. (14)), of DEM wall particles $k$ belonging to a rigid body $p$,:

$$\mathbf{F}_h = -\sum_{k\in\Omega_p} P_k(l^0)^{d-1}\mathbf{n}_k, \tag{30}$$

$$\mathbf{T}_h = -\sum_{k\in\Omega_p} P_k(l^0)^{d-1}\mathbf{r}_{ck} \times \mathbf{n}_k, \tag{31}$$

where $\mathbf{n}_k$ represents the normal vector of the wall at the wall particle $k$, $\mathbf{r}_{ck}$ denotes the position vector between the CM of the rigid body $p$ and the particle $k$, and $\Omega_p$ is the domain with solid wall particles.

The time step of the fluid domain $\Delta t_f$ should follow the Courant–Friedrichs–Lewy (CFL) condition (Courant et al., 1967):

$$\Delta t_f < \frac{l^0 C_r}{|u_{max}|}, \qquad 0 < C_r \leq 1, \tag{32}$$

where $u_{max}$ denotes the maximum velocity and $C_r$ represents the Courant number. Here we adopted the canonical solution of the analytical formulation based on shallow water waves (Ritter, 1892) and assumes the maximum velocity of $|u_{max}| = 2\sqrt{gH_{up}}$. $C_r = 0.2$ is used for all simulations.

Since the transient response for the rigid body contact must be captured in a smaller time step, a dynamic sub-cycling algorithm for rigid bodies is adopted, improving the computational efficiency.

Considering that at least one hundred time steps are used during the solid contact, and taken into account the duration $t_c$ of a typical contact among rigid bodies, based on Hertz's contact theory (Johnson, 1985), an adaptive time step for the solid domain is adopted $\Delta t_s = N_{dt} \Delta t_f$, where $N_{dt}$ denotes an integer, since that:

$$\Delta t_s \leq \min\left(\frac{t_c}{100}\right) \leq \min\left(\frac{2.87}{100}\left[\frac{m_{kl}^2}{l_{kl}^0 E_{kl}^2 \|\mathbf{u}_k - \mathbf{u}_l\|}\right]^{\frac{1}{5}}\right). \tag{33}$$

During one time step $\Delta t_f$, the hydrodynamic effects are assumed to be unchanged and the values of $\mathbf{F}_h$, $\mathbf{F}_g$ and $\mathbf{T}_h$ remain constant. The values are used in the governing equations of solid motion (Eqs. (3) and (4)) during the sub-cycling. On the other hand, contact forces and torques, $\mathbf{F}_c$ and $\mathbf{T}_c$, are computed and considered in Eqs. (3) and (4) at each iteration of the sub-cycling, and then the velocity and position of solid are updated.

In the absence of contact between rigid solids, $\Delta t_s = \Delta t_f$, the solid motion is update considering only the influence of $\mathbf{F}_h$, $\mathbf{F}_g$ and $\mathbf{T}_h$, and the DEM calculations are not carried out.

From the viewpoint of the fluid, the new positions of the DEM particles (wall and internal) are used as the boundary conditions of the MPS, completing the coupling between solid and fluid. A schematic diagram of the algorithm implemented is illustrated in Fig. 6.

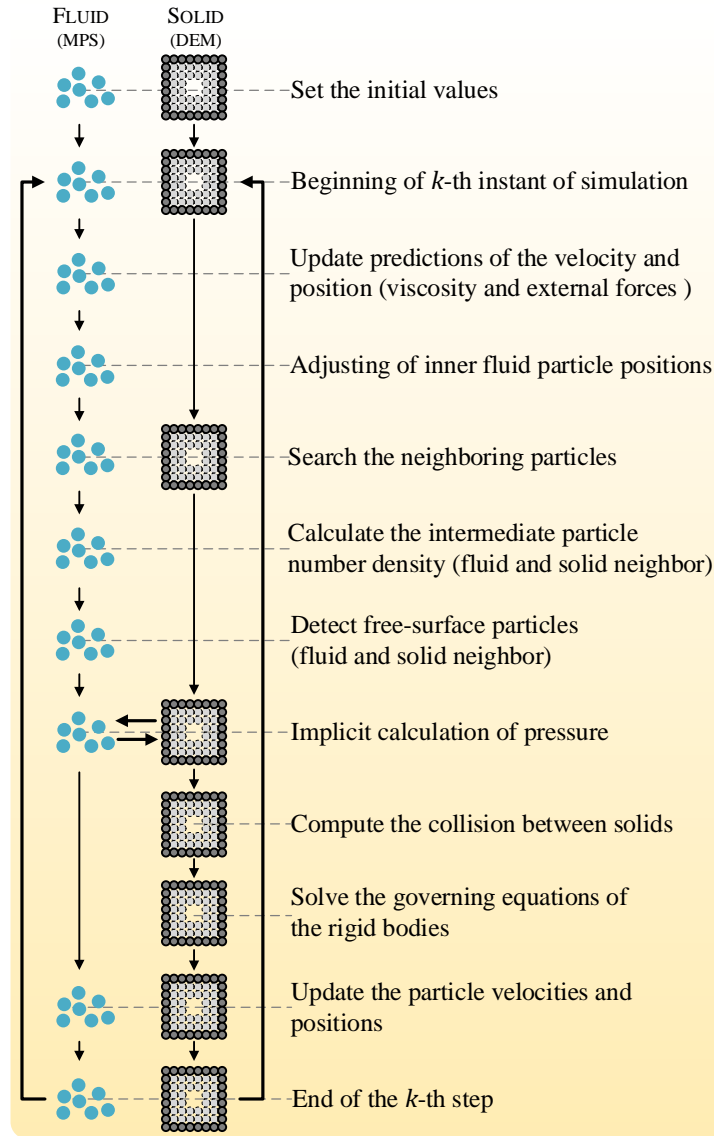

Fig. 6. Schematic diagram of the DEM and MPS coupling algorithm.

## 4.5    *Computational performance*

Particle-based methods suffer from the high computational cost associated with the large number of particles necessary to solve large-scale practical engineering problems. Parallel computing in shared memory systems using OpenMP® 3.0 (Dagum & Menon, 1998) was used to simulate the present cases in reasonable processing time. The numerical simulations with about 1.5 million particles (fluid and solid walls using ~$10^6$ particles and free rigid bodies, i.e., ice floes, using ~$10^3$ particles) for a physical time of 3.0 seconds took approximately 30 to 60 hours with a central processing unit (CPU) Intel® Xeon® Processor E5 v2 Family, a processor base frequency of 2.80 GHz, 20 cores, and 126 GB of memory.

To overcome the high computational cost involving in the simulations of real-world relevant phenomena like ice jam formation and breakup, our next task will be to include the present DEM-MPS model in the 3D parallelized version of our MPS code based on the message passing interface (MPI) (Gropp et al., 1996) strategy and implemented by Fernandes et al. (2015). By using a high-performance computing (HPC) technique, consisting of a combination of a nongeometric dynamic domain decomposition strategy based on particle renumbering and a distributed parallel sorting algorithm for the particle renumbering, the parallelized code enables hybrid parallel processing of hundreds of millions of particles within reasonable runtime in computer CPU clusters.

## 5   Results and discussion

In order to illustrate the ability of the proposed model to numerically reproduce the complex dynamic behavior of the ice-wave interaction phenomenon, it was applied to solid-fluid dam-break flow over dry and wet beds, previously detailed in section 2. For all simulations, we adopted the gravity acceleration $g = 9.81$ m/s². According to the experimental results shown in Fig. 4, the constant vertical velocity of $v_g = 0.4$ m/s was assigned to the gate. The physical properties of the fluid and solids are summarized in Table 2 and Table 3, respectively, and simulation parameters are presented in Table 4. Based on experience of the authors and others, some numerical parameters that can provide reliable predictions were adopted, as illustrated in Table 4. The reader interested in the calibration process of such parameters is invited to refer to Koshizuka and Oka (1996) and Duan et al. (2019) for effective radius $r_e$, Tsukamoto (2016) for surface thresholds $\beta$ and $\varrho$, Lee et al. (2011) for the collision distance $\alpha_1$ and coefficient of restitution $\alpha_2$, and Tsukamoto (2020) for the relaxation coefficient $\gamma$. Adequate damping should be high enough to obtain a non-oscillatory motion of the rigid solids while not too high so that too much energy is dissipated during the collision. Following Amaro Jr. et al. (2019), the damping ratio of the collision $\xi_n = 0.05$ provided a good compromise between static response and low energy dissipation for a complex problem of dam break with multiple blocks and was also adopted herein.

Table 2. Physical properties of the fluid (MPS).

| Property | Water |
|---|---|
| Density $\rho_f$ (kg/m³) | 1000 |
| Kinematic viscosity $v_f$ (m²/s) | $10^{-6}$ |

Table 3. Physical properties of the solids (DEM).

| Property | Tank (Plexiglass) | Block (Polypropylene) |
|---|---|---|
| Density $\rho_S$ (kg/m³) | $\infty$ | 868 |
| Young's modulus $E_s$ (GPa) | 1.0 | 3.3 |
| Poisson's ratio $\upsilon_s$ | 0.37 | 0.40 |
| Static friction $\mu_s$ | 0.412 | 0.412 |

Table 4. Simulation parameters of the fluid (MPS) and solid (DEM) domains.

| Parameter | Value | Parameter | Value |
|---|---|---|---|
| Particle distance $l^0$ (m) | 0.002 | Collision distance $\alpha_1$ | 0.8 |
| Time step (fluid) $\Delta t_f$ (s) | $1.25 \times 10^{-4}$ | Coefficient of restitution $\alpha_2$ | 0.2 |
| Effective radius $r_e$ (m) | $2.1 \times l^0$ | Relaxation coefficient $\gamma$ | 0.01 |
| Surface threshold $\beta$ | 0.97 | Compressibility $\alpha_c$ (ms²/kg) | $10^{-8}$ |
| Surface threshold $\varrho$ | 0.2 | Damping ratio of the collision $\xi_n$ | 0.05 |
| Courant number $C_r$ | 0.2 | | |

### 5.1 Dam breaking with 4 blocks

The free-surface profile evolution of the present numerical model and experimental results are compared for the cases with 4 blocks in dry (Figs. 7($a_1$)-($a_3$)) and wet beds with $H_{do}$ = 1.0 (Figs. 7($b_1$)-($b_3$)), 2.5 (Figs. 7($c_1$)-($c_3$)), and 5.0 (Figs. 7($d_1$)-($d_3$)) cm. It is worth mentioning that the experimental results are from the front camera and exclude the splash. In this sense, splashed fluid particles are omitted in the numerical results. Also, for a cleaner visualization, the blocks are omitted. At the instant $t$ = 0.41 s, the wavefront travels smoothly along the dry bed and hits the downstream wall (Fig. 7($a_1$)). However, the behavior of the flow changes remarkably with the presence of the initial downstream fluid layer. As shown in Figs. 7($b_1$), ($c_1$), and ($d_1$), the downstream fluid remains unaffected by the wave and, consequently, a mushroom-like jet is formed as previously reported by Stansby et al. (1998) and Jánosi et al. (2004). For instance, for $H_{do}$ = 2.5 and 5.0 cm a collapse breaker occurs. As the dam-break flow proceeds, at $t$ = 0.83 s, a backward wave is generated for the cases of dry bed (Fig. 7($a_2$)) and fluid layer of 1 cm depth (Fig. 7 ($b_2$)). At the same time, the wave barely impacts the downstream wall for the downstream fluid depth of 2.5 cm (Fig. 7($c_2$)) and the fluid run-up on the downstream wall occurs for the depth of 5 cm (Fig. 7($d_2$)). Afterwards, at $t$ = 1.50 s, the wave moves downstream after it hits the upstream wall for the dry bed case (Fig. 7($a_3$)), the wave impacts the upstream wall and a portion of the flow forms a vertical run-up jet for $H_{do}$ = 1 and 2.5 cm (Figs. 7($b_2$) and ($c_2$)), and the wave propagates upstream for the fluid layer of 5 cm (Fig. 7($d_2$)). Moreover, during the progress of the fluid flow, the sequence of events such as wavefront progress, wave impact, and run-up is delayed as the downstream fluid layer increases.

Comparing the free-surface profile between the numerical and the experimental results shows that the present numerical model has successfully reproduced the dam-break flow behavior for all downstream fluid depths. However, Figs. 7(c₁) and (d₁) show small deviations between experimental and numerical wave events computed by the present model. Such discrepancy can be seen in other particle-based simulations (Crespo et al., 2008; Farzin et al., 2019; Soleimani & Ketabdari, 2020; Ye et al., 2020). As stated in Duan et al. (2019), stabilization adjustments usually adopted by particle-based methods inevitably bring some numerical dissipation. Here, the first-order pressure gradient, see Eq. (18), used to prevent numerical instability, has a particle stabilizing term (PST), i.e., an adjustment technique represented by the adoption of the minimum pressure $\hat{P}_i$. Therefore, this numerical dissipation associated with variations in the gate removal process might be possible reasons for such discrepancy.

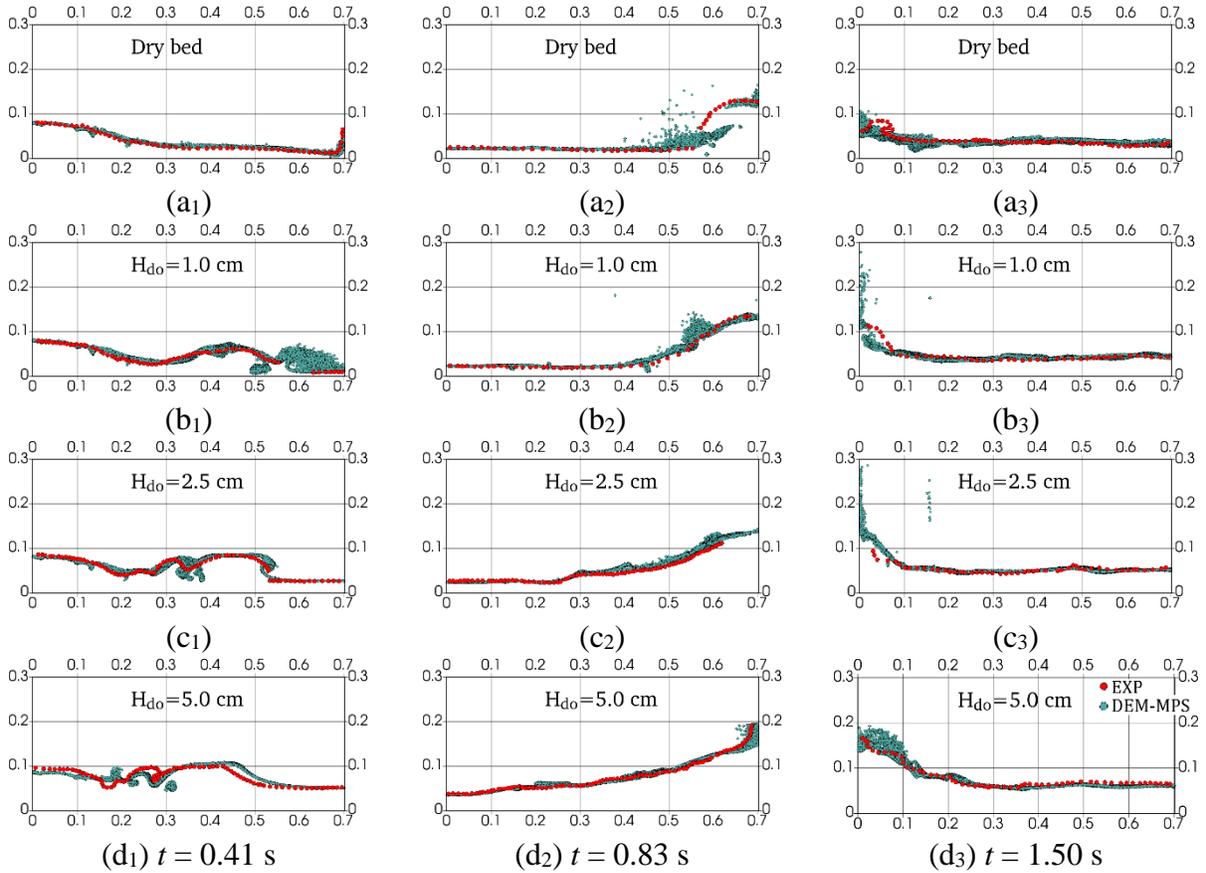

Fig. 7. Evolution of the experimental (red dots) and numerical (turquoise dots) wave profiles for dam breaking with 4 blocks at the instants $t$ = 0.41, 0.83, 1.50 s (from left to right). (a₁, a₂, a₃) Dry bed and wet beds $H_{do}$ = (b₁, b₂, b₃) 1.0, (c₁, c₂, c₃) 2.5 and (d₁, d₂, d₃) 5.0 cm. Experimental and numerical results are from the front camera and exclude the splash.

The snapshots of the dam breaking with 4 blocks from the experiments and simulations are presented in Figs. 8-11 (see also Video 1). The top view of the initial position of the 4 blocks is highlighted at the top left. Delaunay triangulation filter of the open-source Paraview (Ahrens et al., 2005) is used to represent the blocks as a mesh of triangle polygons. The dry bed case is shown in Fig. 8. After the gate is released, the blocks are transported downstream. The experimental and numerical results show similar behaviors of blocks, nevertheless, the computed evolution of blocks A1 and B1 are slightly faster than those measured in the experiments until $t = 0.8$ s. After that, a backward wave transports the blocks back upstream until the instant $t = 1.6$ s. The experimentally measured and numerically computed positions of the A2 block are in good agreement, but the remaining blocks' positions show some differences. Such discrepancies are expected considering the challenge of spacing evenly the blocks initially and the chaotic features of the phenomena through time, such as breaking waves, wave-splash, and plunging jet. The blocks are transported downstream after $t = 2.0$ s, and the flow starts to calm down. As we will show later, chaotic flow features after the wave hits the downstream wall may not be completely reproduced in experimental repetitions.

The dam breaking with the 1 cm depth fluid layer is depicted in Fig. 9. During the first 2.4 s, where the blocks were transported by the fluid, very good agreement was obtained between experimental and numerical solid positions.

In Fig. 10, experimental and numerical solid positions for the fluid layer of 2.5 cm present a good agreement until $t = 1.6$ s. Nevertheless, after $t = 2.0$ s, the blocks A2 and B2 are transported downstream in the experiment while very slow motions are numerically computed.

Finally, the dam breaking with the 5 cm depth fluid layer is illustrated in Fig. 11. A good agreement between experimental and numerical solid motions can be seen until the instant $t = 1.2$ s. After $t = 1.2$ s, where the blocks move upstream and collisions among the blocks occur, the computed positions of the blocks A1 and B1 remain close to the upstream wall of the tank in the numerical results while these blocks are carried by the flow in the experimental results. In summary, numerical and experimental solid motions show qualitatively identical features.

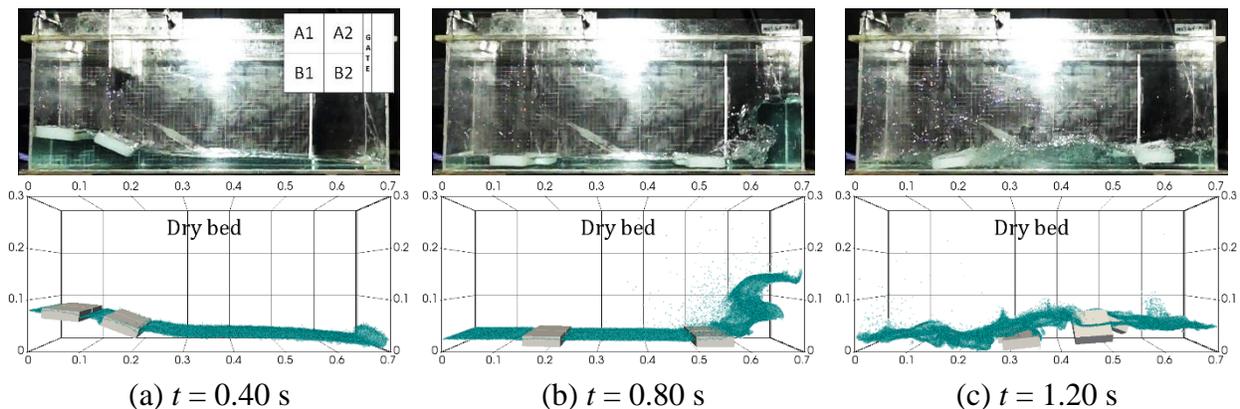

(a) $t = 0.40$ s        (b) $t = 0.80$ s        (c) $t = 1.20$ s

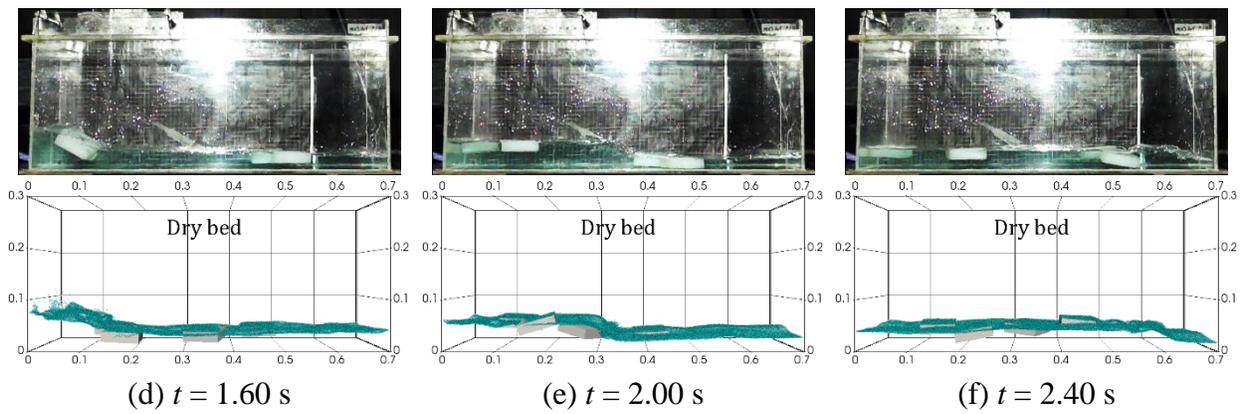

(d) $t$ = 1.60 s       (e) $t$ = 2.00 s       (f) $t$ = 2.40 s

Fig. 8. Snapshots of the experimental and numerical dam breaking with 4 blocks and dry bed at the instants $t$ = 0.40, 0.80, 1.20, 1.60, 2.00, 2.40 s (front view).

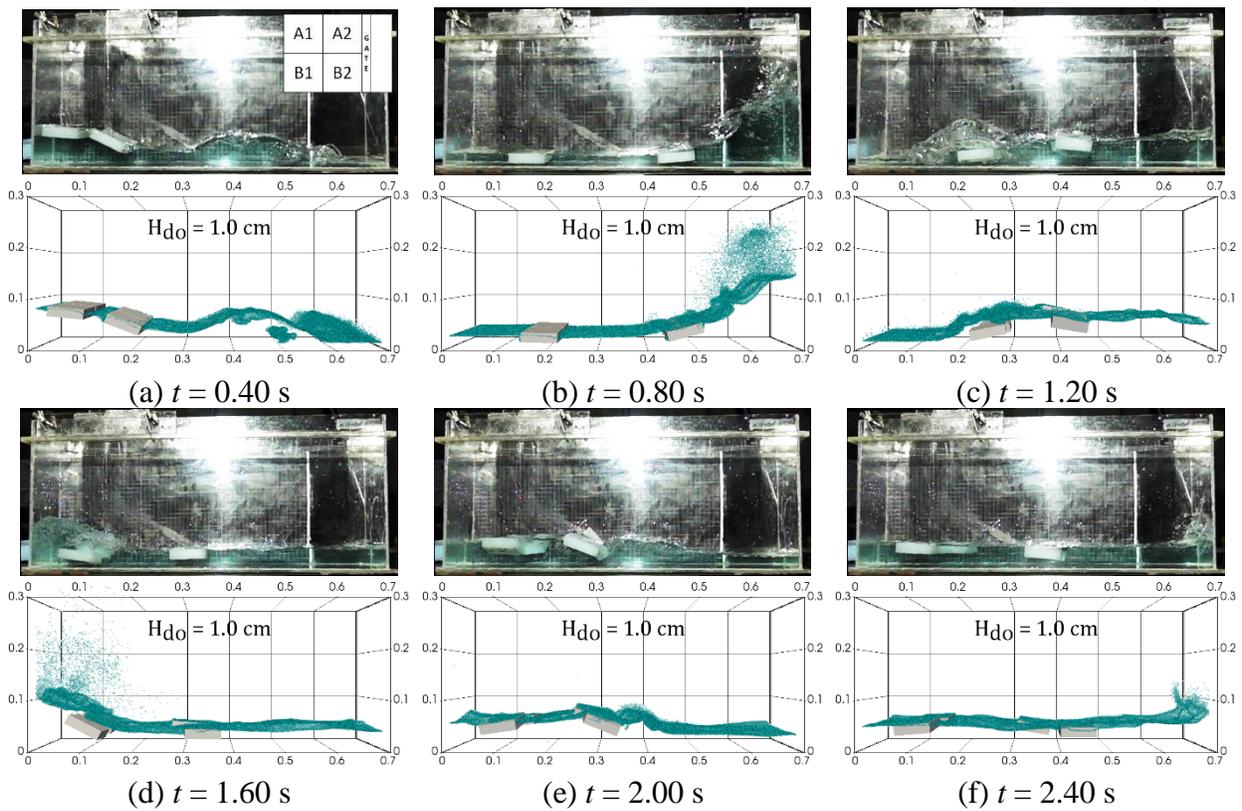

(a) $t$ = 0.40 s       (b) $t$ = 0.80 s       (c) $t$ = 1.20 s

(d) $t$ = 1.60 s       (e) $t$ = 2.00 s       (f) $t$ = 2.40 s

Fig. 9. Snapshots of the experimental and numerical dam breaking with 4 blocks and wet bed $H_{do}$ = 1 cm at the instants $t$ = 0.40, 0.80, 1.20, 1.60, 2.00, 2.40 s (front view).

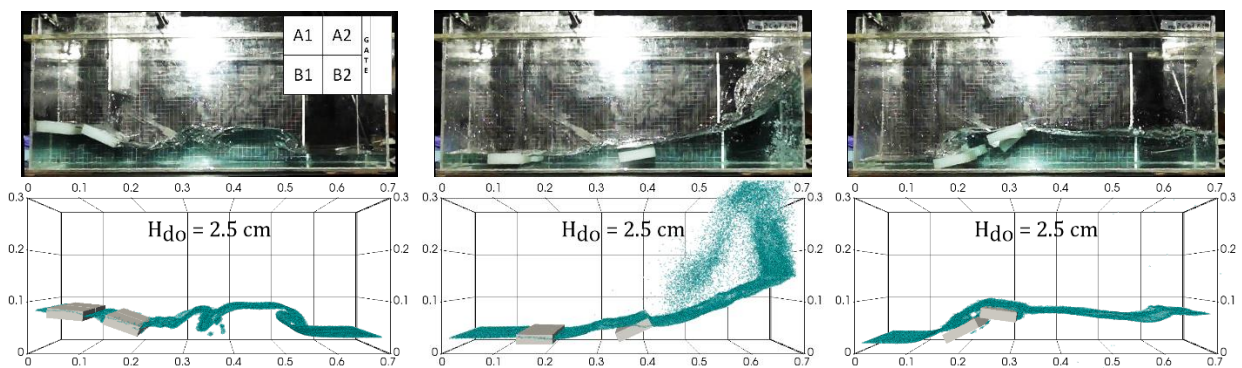

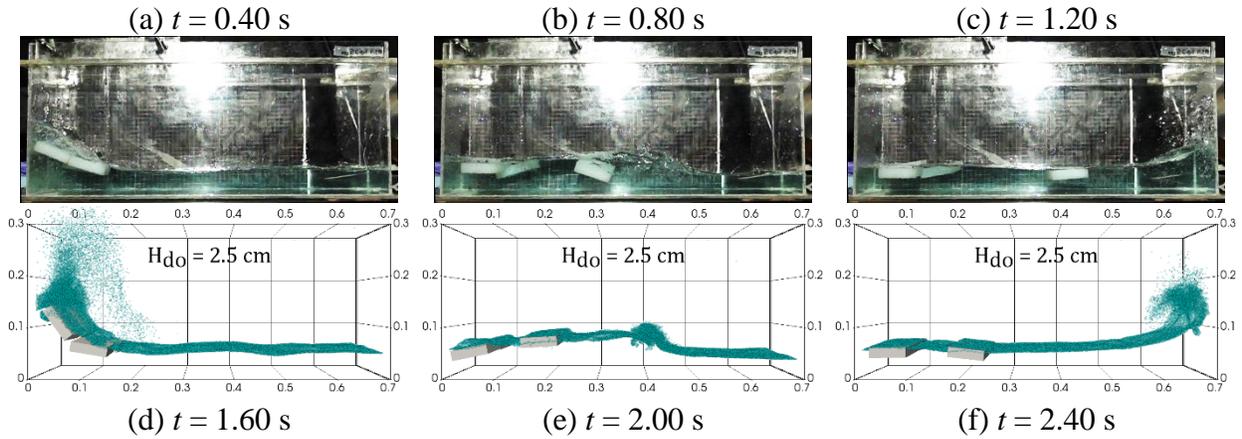

Fig. 10. Snapshots of the experimental and numerical dam breaking with 4 blocks and wet bed $H_{do}$ = 2.5 cm at the instants $t$ = 0.40, 0.80, 1.20, 1.60, 2.00, 2.40 s (front view).

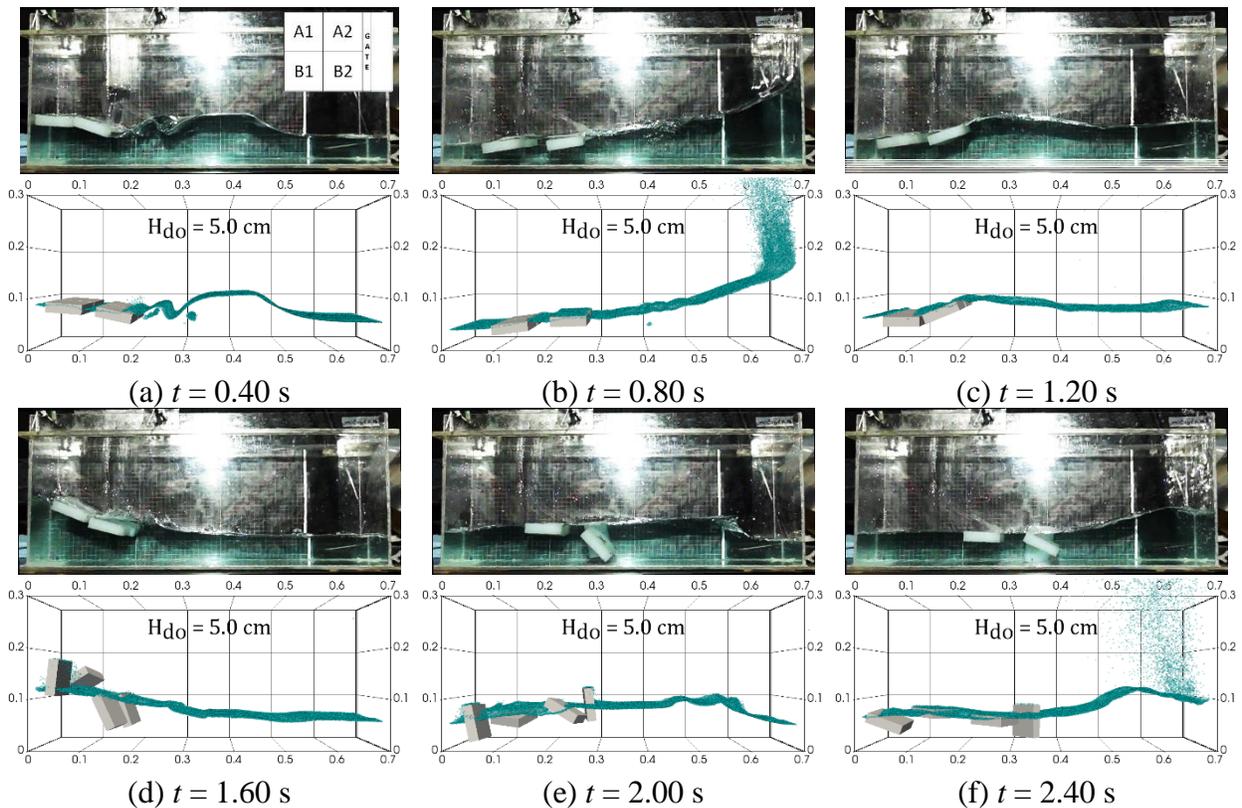

Fig. 11. Snapshots of the experimental and numerical dam breaking with 4 blocks and wet bed $H_{do}$ = 5 cm at the instants $t$ = 0.40, 0.80, 1.20, 1.60, 2.00, 2.40 s (front view).

The longitudinal and vertical motions of each block have been plotted to quantify the comparisons. Figures 12(a) and 12(b) show the experimental and numerical longitudinal motions $X_{B1}$ and $X_{B2}$ of blocks B1 and B2, respectively, for the dry bed case during the first 3 s after the gate removal. Here the three experimental repeats have been named EXP 1, EXP 2, and EXP 3. As Fig. 12(a) shows, the initial trajectory of block B1 obtained numerically follows the tendency of the experimental one, but eventually, the block is transported downstream faster numerically than experimentally.

While the maximum distance reached by the block B1 at $t = 1.5$ s is about 0.18 m in the experiments, it was overestimated at a distance of 0.3 m in the numerical simulation. The main reason for this discrepancy might be the difficulty in setting the initial arrangement of the blocks in regular and equally spaced positions, which might lead to values different from those numerically computed. In addition to this, the variations in the gate removal process might also influence the motions. After $t = 1.5$ s, block B1 is carried by the reflected wavefront, and larger discrepancies between the numerical and experimental results occur. These discrepancies, which also occur between the experimental runs, are reasonable due to the chaotic nature at this stage, because of an intense splash with air-water mixture created by the fall of the vertical run-up jet onto the underlying fluid. In fact, the breaking processes (e.g., wave-splash, plunging jet penetration depth, and breaking wave) exhibit chaotic behaviors, as previously revealed by the extensive investigation of laboratory observations and numerical simulations in Wei et al. (2018). On the other hand, the longitudinal motions of the block B2 (Fig. 12(b)) are correctly reproduced by the numerical simulations until $t = 1.0$ s. After that, the computed motion presents the same tendency as the experimental ones.

As shown in Fig. 13, the measured vertical motions $Y_{B1}$ and $Y_{B2}$ of blocks B1 and B2, respectively, were accurately predicted by the numerical simulations, indicating that the buoyancy force is well reproduced by the numerical model.

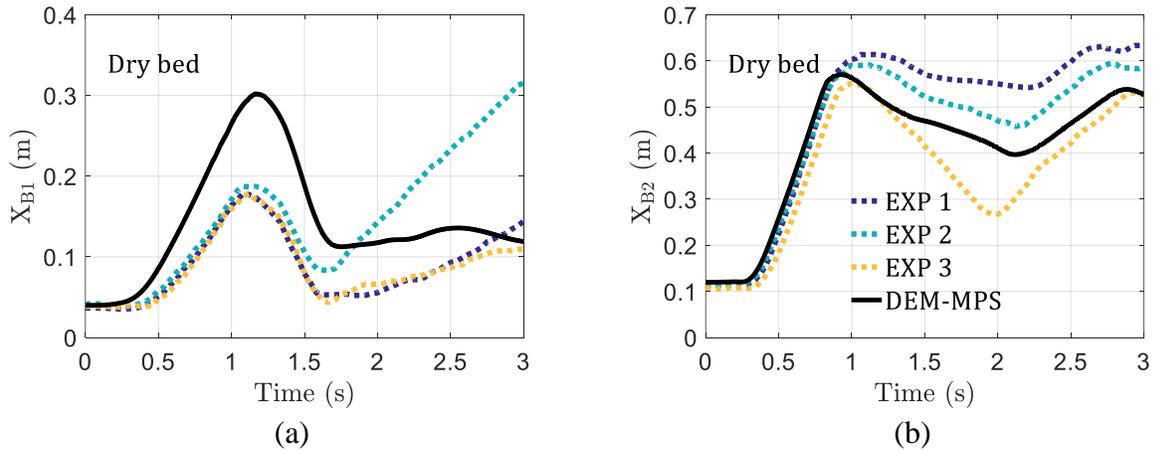

Fig. 12. Dam breaking with 4 blocks and dry bed. Experimental (dotted lines) and numerical (solid line) motions of the (a) block B1 and (b) block B2 along the longitudinal direction.

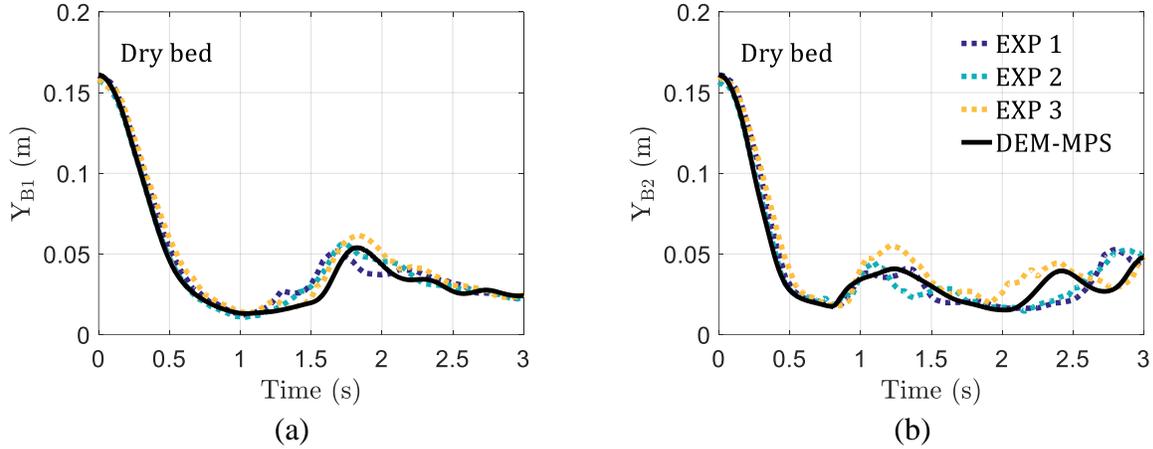

(a)                                  (b)

Fig. 13. Dam breaking with 4 blocks and dry bed. Experimental (dotted lines) and numerical (solid line) motions of the (a) block B1 and (b) block B2 along the vertical direction.

Figure 14 shows the experimental and numerical longitudinal motions $X_{B1}$ and $X_{B2}$ of blocks B1 and B2, respectively, in dam-break flows with downstream depths of $H_{do}$ = 1, 2.5, and 5 cm. For the depth of $H_{do}$ = 1 cm (Fig. 14(a$_1$)), the block B1 is transported downstream by the wave front, reaching a maximum distance between 0.20 and 0.26 m at around $t$ = 1.2 s in the experiments, whereas in the numerical simulation block B1 moves slightly faster and reaches the maximum distance of 0.28 m at around $t$ = 1.1 s. Subsequently, the block B1's experimental and numerical motions are in good agreement. After $t$ = 1.5 s, chaotic features similar to the dry bed case occur due to the violent hydrodynamic impact followed by the wave breaking, and the computed motion matches the measured motions from the second experimental run only. As shown in Fig. 14(b$_1$), the measured motions of block B2 are very well predicted by the numerical model. For the depth of $H_{do}$ = 2.5 cm (Fig. 14(a$_2$)), a maximum distance between 0.18 and 0.26 m at around $t$ = 1.25 s is reached by block B1 in the experiments, whereas block B1 moves slightly faster in the numerical simulation, reaching a maximum distance of 0.23 m at around $t$ = 1.1 s. During the upstream return of block B1, the computed motion is in reasonable agreement with the experimental ones. Again, due to the chaotic features of the breaking wave process after $t$ = 1.5 s, the experimental motions present distinct patterns between them, while a slower motion was numerically obtained, and the block remains almost in the same position. The measured and computed motions of block B2 are in very good agreement until $t$ = 1.0 s, but, subsequently, the computed motion matches well the results from the first experiment while underestimating the results from the second and third experiments (Fig. 14(b$_2$)). Finally, for the higher bed depth $H_{do}$ = 5 cm (Fig. 14(a$_3$)), the block B1's maximum displacements measured experimentally vary between 0.08 and 0.11 m while the computed maximum distance reaches 0.12 m, both occurring at around $t$ = 0.8 s. When the block

B1 moves back upstream, the computed motion is in reasonable agreement with the first experimental results. After $t = 1.5$ s, significant discrepancies between numerical and experimental motions of block B1 were obtained. The numerical and experimental results prior to $t = 1.5$ s show that the reflected wave transported the blocks A2 and B2 towards the blocks A1 and B1, which collide against the upstream wall. Afterwards, the block B1 was submerged into water and was transported by the wave. Nevertheless, in the experimental runs, the motion of block B1 is more affected by the thrust generated by the wave than in the numerical simulation. A good agreement is obtained between experimental and numerical motions for the block B2, despite the underestimation of the block's displacement in the numerical simulation after $t = 1.9$ s, as shown in Fig. 14(b$_3$).

The vertical motions $Y_{B1}$ and $Y_{B2}$ of blocks B1 and B2, respectively, are provided in Figs. 15(a$_1$)-(a$_3$) and (b$_1$)-(b$_3$). Although some minor discrepancies are observed, the trends of the computed vertical motions of both blocks agree well with the experimental ones.

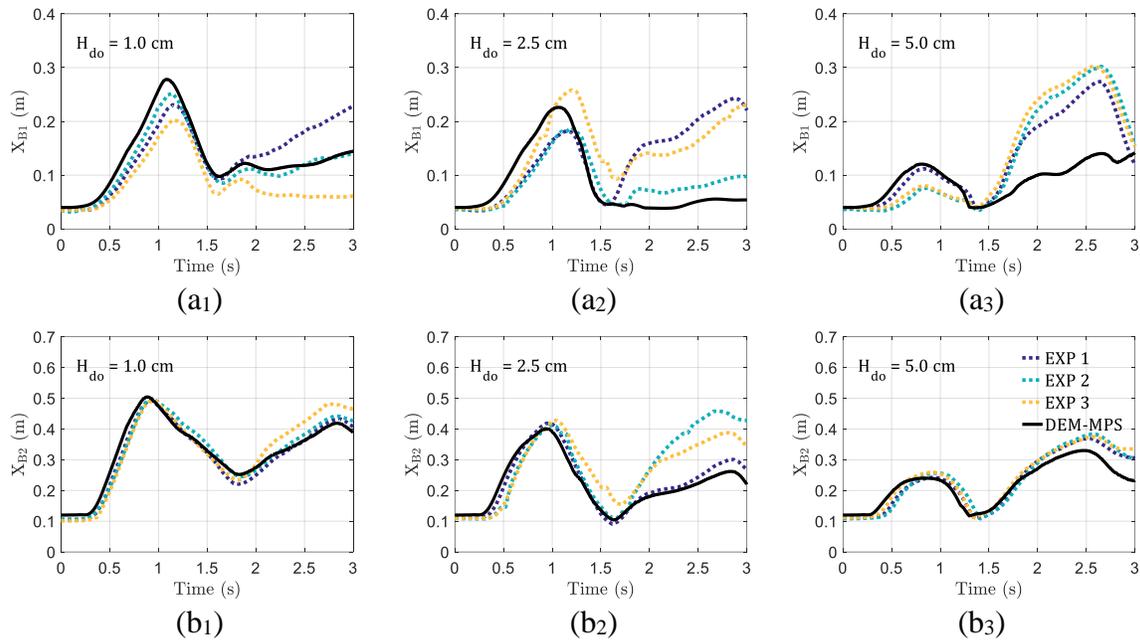

Fig. 14. Dam breaking with 4 blocks and wet beds: $H_{do}$ = 1, 2.5 and 5 cm. Experimental (dotted lines) and numerical (solid line) motions of the (a$_1$, a$_2$, a$_3$) block B1 and (b$_1$, b$_2$, b$_3$) block B2 along the longitudinal direction.

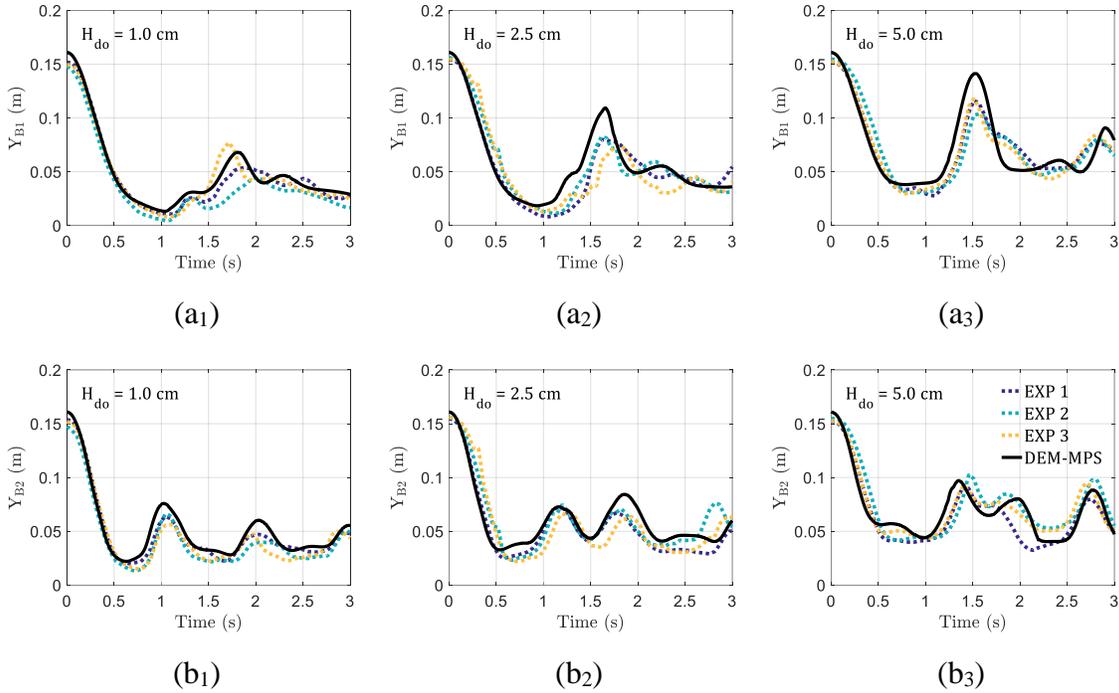

Fig. 15. Dam breaking with 4 blocks and wet beds: $H_{do}$ = 1, 2.5 and 5 cm. Experimental (dotted lines) and numerical (solid line) motions of the ($a_1$, $a_2$, $a_3$) block B1 and ($b_1$, $b_2$, $b_3$) block B2 along the vertical direction.

The maximum longitudinal displacements $X_{A1,max}$, $X_{A2,max}$, $X_{B1,max}$ and $X_{B2,max}$ reached by the blocks A1, A2, B1, and B2, respectively, from the instant $t = 0$ to $t = 1.5$ s are illustrated in Fig. 16. Since the initial potential energy decreases with the increase of initial fluid depth in the bed ($H_{do}$) and the initial energy dissipation is higher for wet beds due to the plunging or collapse wave breaking caused by the hydraulic jump, it is expected that the increase of $H_{do}$ will lead to a decrease of the wave front speed and, as a consequence, a decrease on the maximum distance reached by the blocks during the dam-break phase. However, the results provided by the experiments do not match exactly this behavior, and some variations can be observed between $H_{do} = 0$ and 1 cm for the blocks A1 and B1, initially placed closer to the upstream tank wall, as shown in Figs. 16(a) and (c). On the other hand, the numerical results for these blocks presented the anticipated behavior. The experimental and numerical results for the blocks A2 and B2 are in very good agreement with the expected behavior.

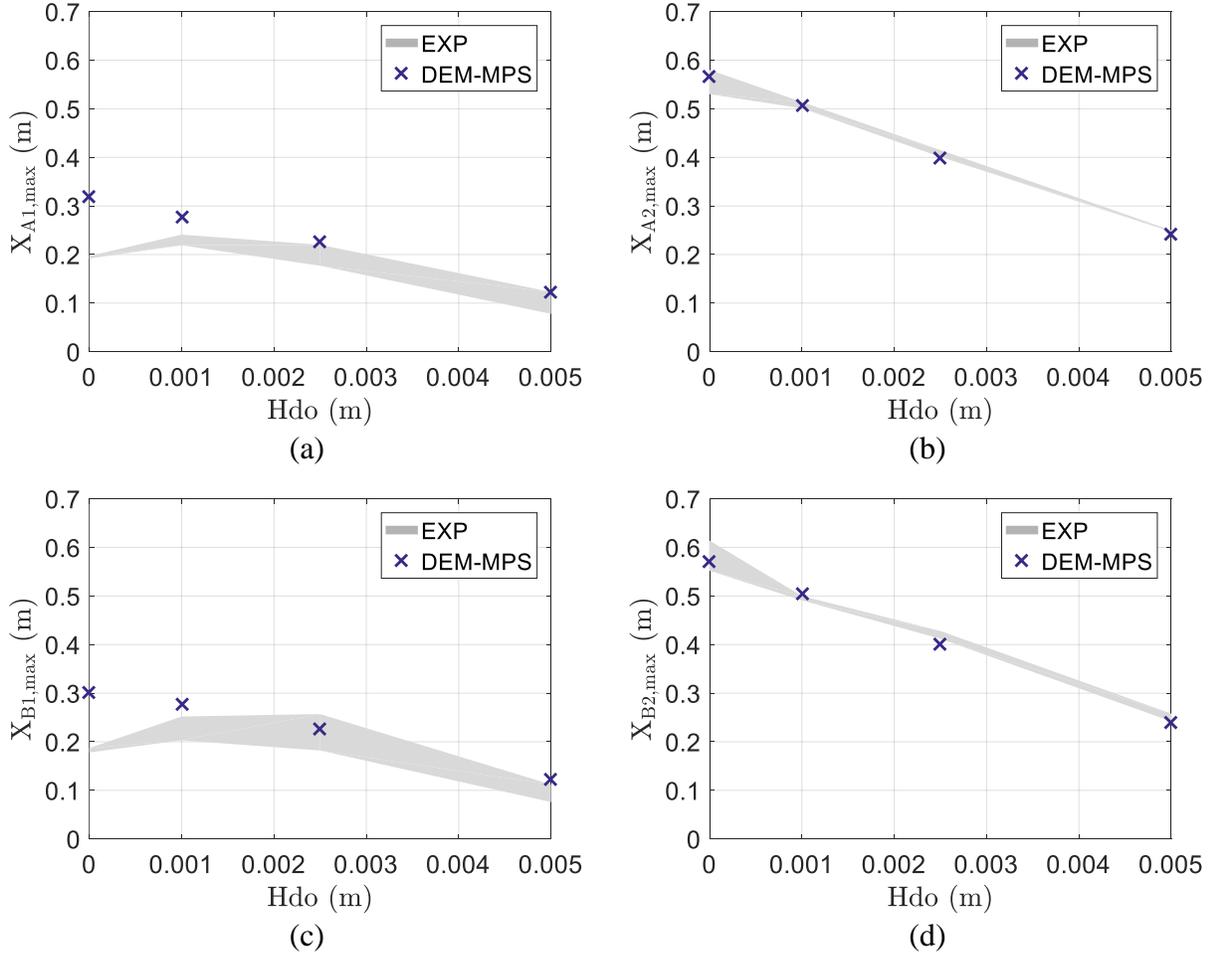

Fig. 16. Dam breaking with 4 blocks over dry and wet beds. Maximum longitudinal displacement of blocks (a) A1, (b) A2, (c) B1 and (d) B2. Experimental and numerical results between the instants $t = 0$ and $t = 1.5$ s.

### 5.2   Dam breaking with 9 blocks

The free-surface profiles of the present model and experimental runs for the cases with 9 blocks in dry and wet beds are compared in Fig. 17. Again, the splash is not presented for the wave profiles, and the blocks are omitted. At the early stage, $t = 0.41$ s, the formation of the mushroom-like jet occurs when a downstream fluid layer is present (Fig. 17($b_1$), ($c_1$) and ($d_1$)). Furthermore, as previously discussed for the cases with 4 blocks (see Fig. 7), at the instants $t = 0.83$ s and $t = 1.50$ s, the increase of the downstream fluid layer leads to slower wave propagation. In other words, the sequence of events such as wave impact and run-up happens later with the increase of the downstream depth. In general, as in the previous cases with 4 blocks, there is a satisfactory agreement between numerical and experimental free-surface profiles. By comparing the wave profiles of the 9 block case (Fig. 17) with those of the 4 block case (Fig. 7), the differences, despite being relatively small, are more evident in the later stage of the hydrodynamic process. Thus, the presence of the light-weight blocks is almost negligible in the initial gravity dominant dam-break event. However, the fluid-solid interaction is not negligible in the late hydrodynamic impact and wave breaking events. A detailed discussion about the effect of ice floes on the flow is presented in Appendix B.

Figures 18-21 (also see Video 2) show a sequence of frames of the dam breaking with 9 blocks from the experimental repeats and numerical simulations. The top view of the 9 blocks is highlighted at the top left. Fig. 18 shows a reasonable agreement between experimental and numerical block motions until $t$ = 1.2 s. Afterwards, the merging of the collapsed upward water flow and reflected wave dramatically disturbs the water surface and, as a result, a more chaotic motion of the solids is generated, especially for the solids downstream. Consequently, the computed solid positions deviate from the experimental ones after $t$ = 1.6 s. For the dam breaking scenario with $H_{do}$ = 1 cm, experimental solid motions are well predicted by the numerical simulations until $t$ = 1.2 s (Fig. 19). At the instant $t$ = 1.6 s, the solid motions are disturbed by the combination of a reflected wave and the collapsed water flow. As a result, the computed solid motions deviate from the experimental ones, although the present model reproduced the main features of the solid trajectories. Figs. 20 and 21 depict the dam breaking with a fluid layer of 2.5 and 5 cm depth, respectively. The numerical predictions of the solid motions are in good agreement with the experimental ones through all of the 2.4 s considered in the analyses, although they do not match exactly. In general, the results indicated that the present model is able to predict the solid motions with satisfactory details.

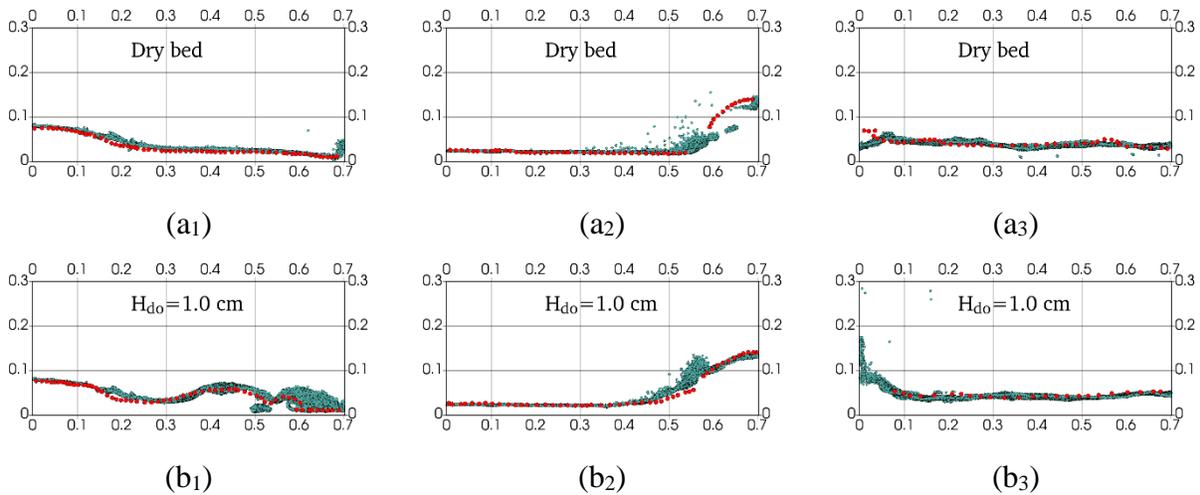

(a₁)          (a₂)          (a₃)

(b₁)          (b₂)          (b₃)

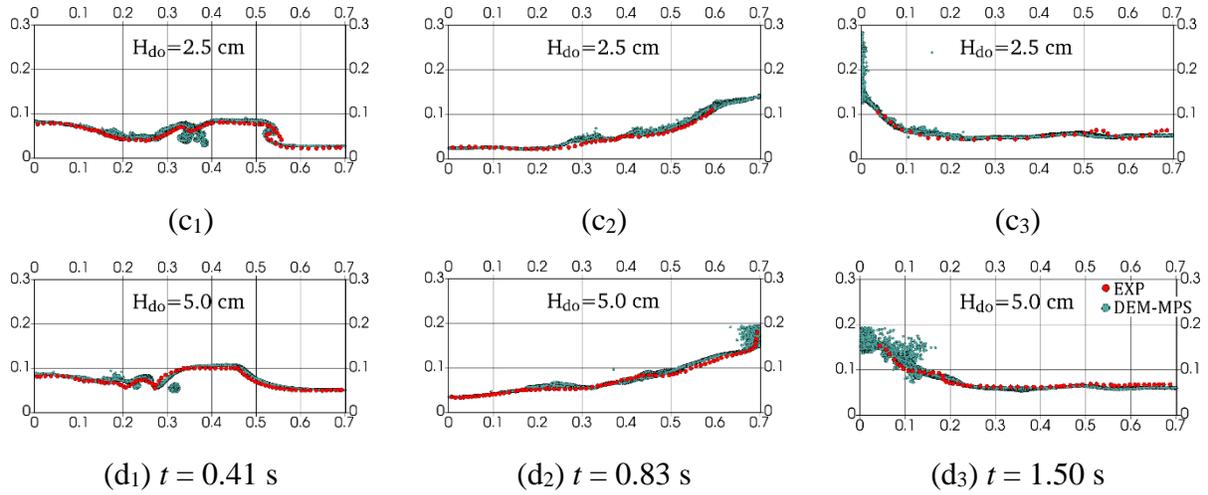

(c₁)                    (c₂)                    (c₃)

(d₁) *t* = 0.41 s        (d₂) *t* = 0.83 s        (d₃) *t* = 1.50 s

Fig. 17. Evolution of the experimental (red dots) and numerical (turquoise dots) wave profiles for dam breaking with 9 blocks at the instants *t* = 0.41, 0.83, 1.50 s (from left to right). (a₁, a₂, a₃) Dry bed and wet beds $H_{do}$ = (b₁, b₂, b₃) 1, (c₁, c₂, c₃) 2.5 and (d₁, d₂, d₃) 5.0 cm. Experimental and numerical results are from the front camera and exclude the splash.

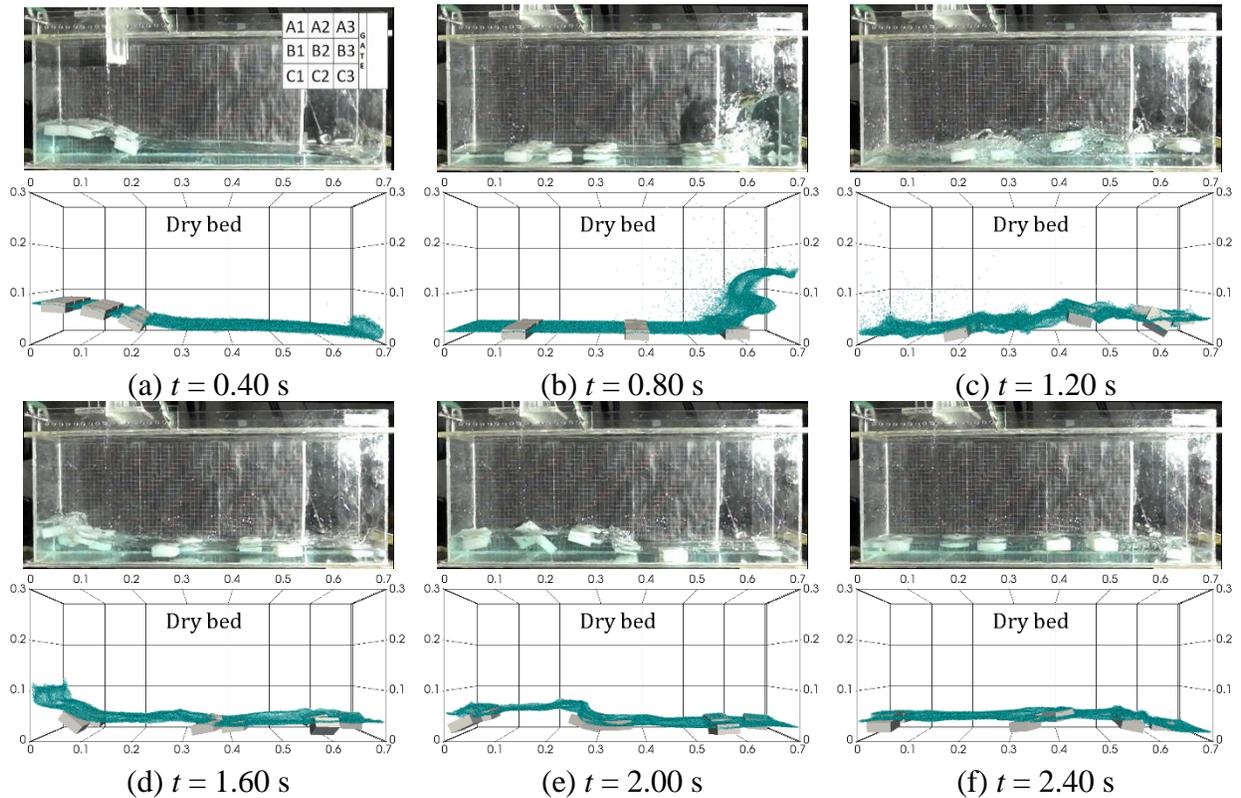

(a) *t* = 0.40 s         (b) *t* = 0.80 s         (c) *t* = 1.20 s

(d) *t* = 1.60 s         (e) *t* = 2.00 s         (f) *t* = 2.40 s

Fig. 18. Snapshots of the experimental and numerical dam breaking with 9 blocks and dry bed at the instants *t* = 0.40, 0.80, 1.20, 1.60, 2.00, 2.40 s (front view).

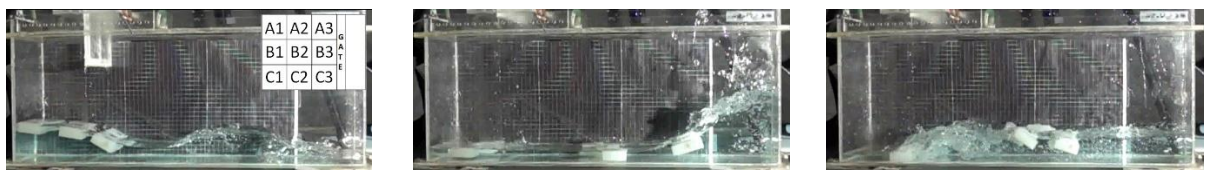

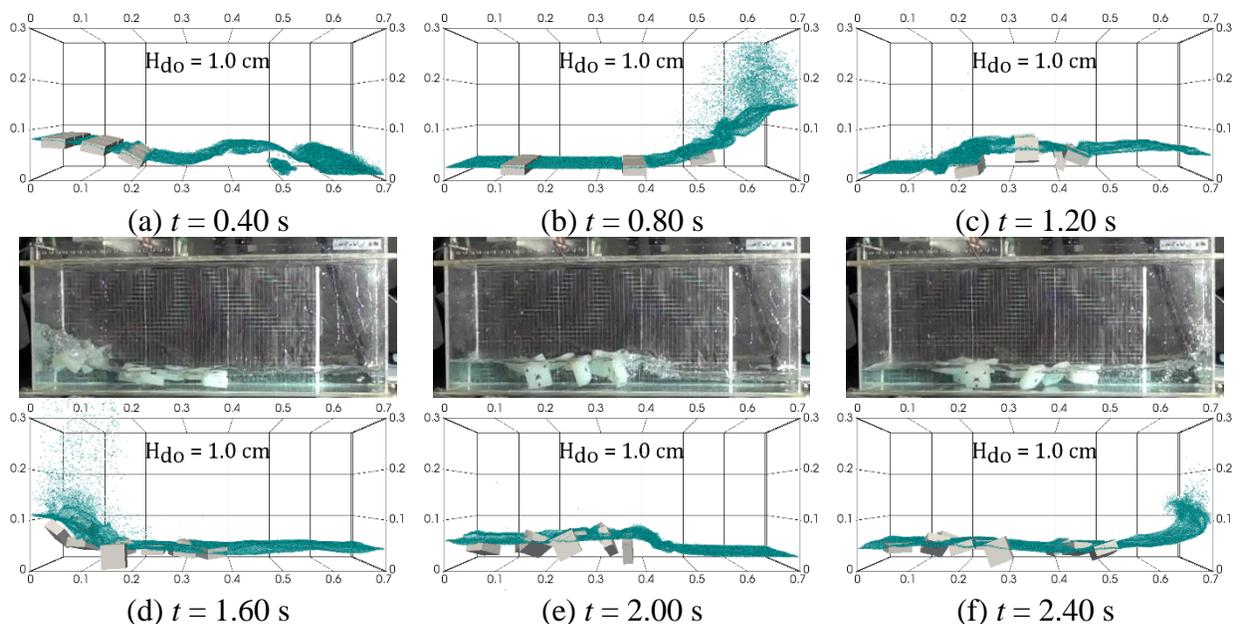

Fig. 19. Snapshots of the experimental and numerical dam breaking with 9 blocks and wet bed $H_{do} = 1$ cm at the instants $t = 0.40$, 0.80, 1.20, 1.60, 2.00, 2.40 s (front view).

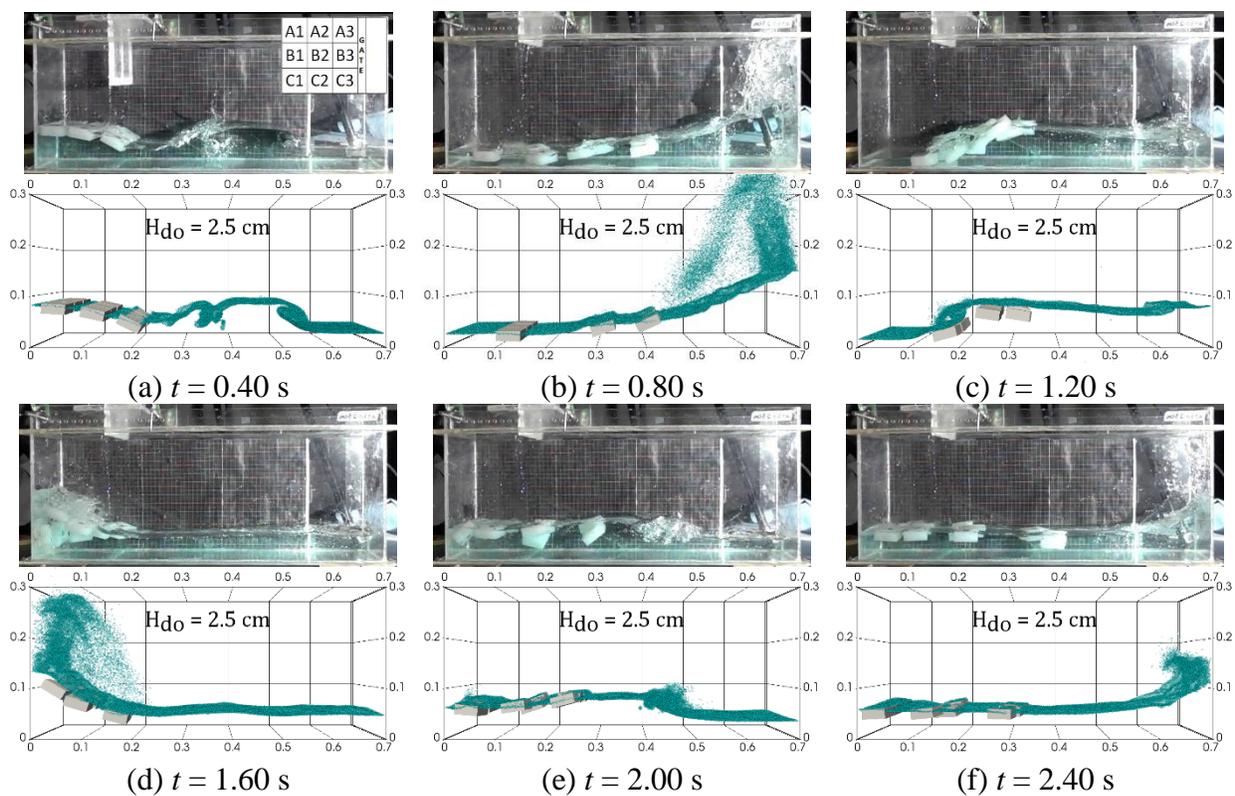

Fig. 20. Snapshots of the experimental and numerical dam breaking with 9 blocks and wet bed $H_{do} = 2.5$ cm at the instants $t = 0.40$, 0.80, 1.20, 1.60, 2.00, 2.40 s (front view).

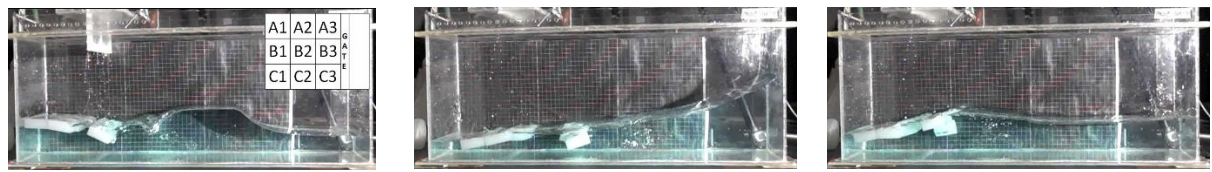

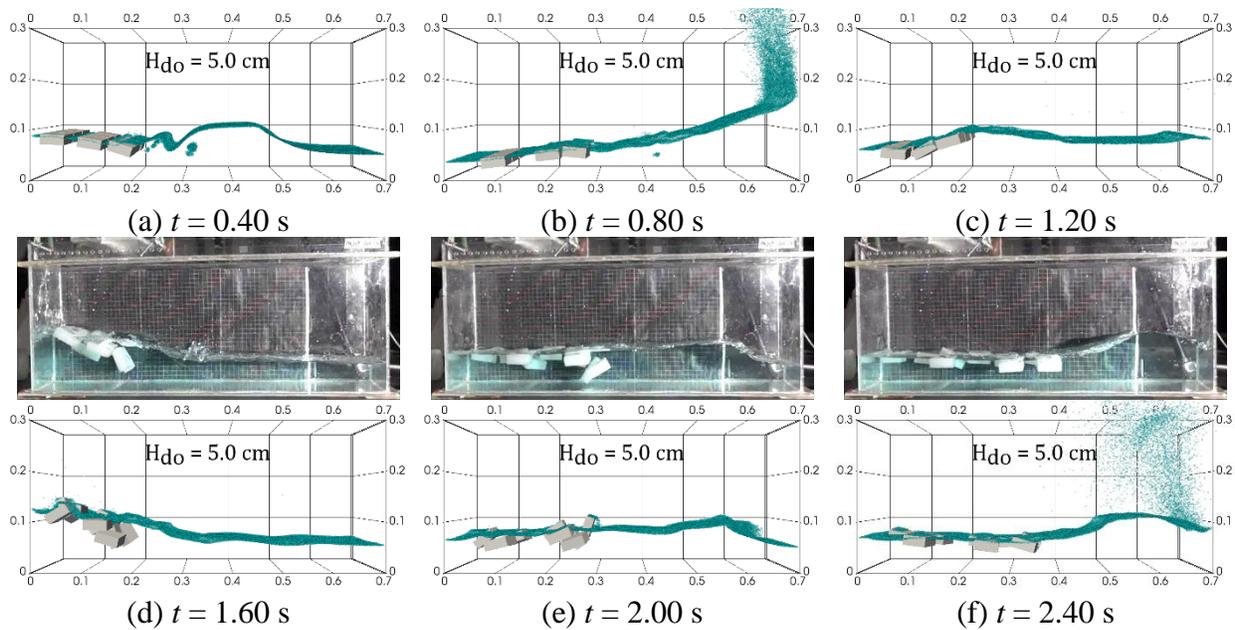

Fig. 21. Snapshots of the experimental and numerical dam breaking with 9 blocks and wet bed $H_{do}$ = 5 cm at the instants $t$ = 0.40, 0.80, 1.20, 1.60, 2.00, 2.40 s (front view).

Fig. 22 provides the experimental and numerical longitudinal motions $X_{C1}$, $X_{C2}$, and $X_{C3}$ of blocks C1, C2, and C3, respectively. The large variability of the maximum displacement in the dam-break phase reached by block C1 in the experiments, ranging from 0.08 to 0.22 m, indicates that its motion is significantly affected by any initial experimental disturbance (Fig. 22(a)). Besides, the numerical motion agrees well with the third experiment but overestimates the results from the first and second experiments until $t$ = 1.7 s, approximately. Subsequently, while in the experiments the block C1 hits the wall, collides with others, and is transported downstream, in the numerical simulations, the block only hits the wall and is transported downstream very slowly, almost remaining in the same position. In addition, as in the previous cases with 4 blocks, the fluid flow becomes chaotic at this stage, and the solid motions are influenced by many factors such as wave breakings and multiple solid collisions. According to the experimental results, the block C2 is transported downstream, reaching the maximum longitudinal motions between 0.33 and 0.43 at around $t$ = 1.2 s, and then moves upstream until $t$ = 1.8 s, approximately (Fig. 22(b)). After that, distinct patterns were obtained from the experimental repeats, and the computed motion presents a tendency similar to the first experiment. Fig. 22(c) shows that during the first 0.9 s, the numerical motions of block C3 match well the experimental motions from EXP 1 and EXP 3, but overestimate EXP 2. Subsequently, the computed motion is closer to the results from EXP 1 and EXP 3.

Fig. 23 compares the experimental and numerical vertical motions $Y_{C1}$, $Y_{C2}$, and $Y_{C3}$ of blocks C1, C2, and C3, respectively. The overall trends of the computed vertical motions agree well with the experimental ones. However, as discussed in the previous section for 4 blocks, some local discrepancies are also present.

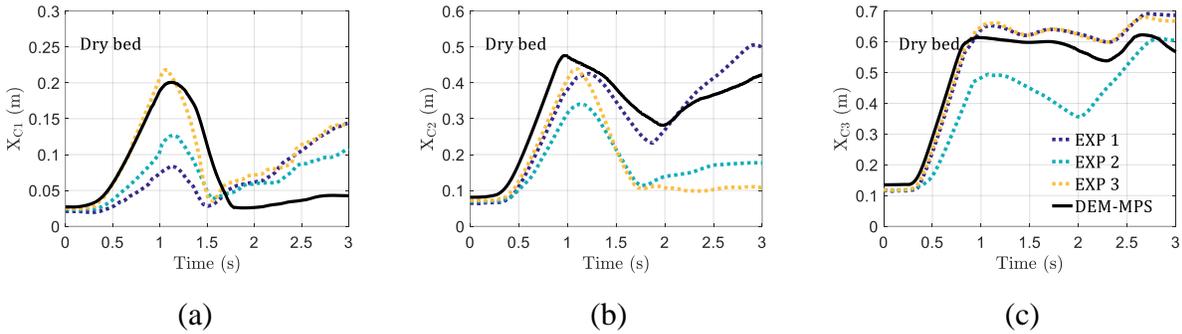

Fig. 22. Dam breaking with 9 blocks and dry bed. Experimental (dotted lines) and numerical (solid line) motions of the (a) block C1, (b) block C2 and (c) block C3 along the longitudinal direction.

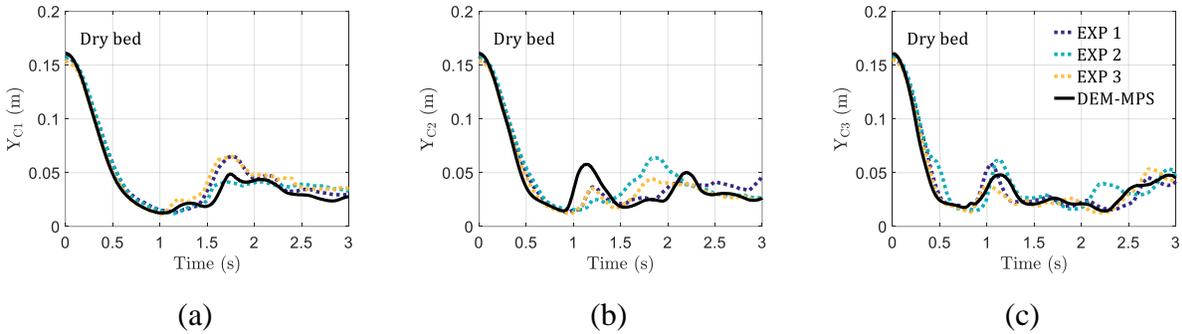

Fig. 23. Dam breaking with 9 blocks and dry bed. Experimental (dotted lines) and numerical (solid line) motions of the (a) block C1, (b) block C2 and (c) block C3 along the vertical direction.

In Fig. 24, the computed longitudinal motions $X_{C1}$, $X_{C2}$, and $X_{C3}$ of blocks C1, C2, and C3, respectively, are compared against the experimental ones, considering the three downstream depths of $H_{do}$ = 1, 2.5, and 5 cm. As in the previous cases with the dry bed, the experiments present a noticeable variation of the initial maximum distances reached by the block C1 for $H_{do}$ = 1.0 and 2.5 cm (Fig. 24($a_1$) and ($a_2$)). Besides, the computed motions for these two depths agree well with the experimental ones, although, for $H_{do}$ = 1 cm, the block C1 moves faster and reaches a maximum distance of about 0.22 m, higher than the maximum distance around 0.18 m measured in the experiments. For $H_{do}$ = 5 cm, the motion of block C1 is well reproduced by the numerical simulations during the first 1.5 s, and after, it matches only the third experiment but underestimates the values from the first and second ones (Fig. 24($a_3$)).

A good agreement between numerical and experimental motions of block C2 for $H_{do}$ = 1 and 2.5 cm during all intervals of 3 s is shown in Fig. 24($b_1$) and ($b_2$). Furthermore, the computed motion is overestimated for $H_{do}$ = 5 cm during the first 1.3 s, approximately, followed by a better agreement between $t$ = 1.3 s and $t$ = 3.0 s (Fig. 24($b_3$)).

For all downstream depths, the motions of block C3 are well reproduced by the numerical simulations (Fig. 24($c_1$)-( $c_3$)).

Fig. 25 shows the experimental and numerical vertical motions $Y_{C1}$, $Y_{C2}$, and $Y_{C3}$ of blocks C1, C2, and C3, respectively. When compared with experimental motions, all computed motions present some local discrepancies, but overall features of the experimental vertical motions are very well reproduced by the numerical model.

The overall good agreement between experimental and computed motions demonstrates that the proposed model is able to reproduce the main behaviors of the complex fluid-solid interaction phenomenon, suggesting that it can be extended for the simulation of real-scale ice-wave interaction problems, such as ice-jam formation and breakup and ice-structure interaction.

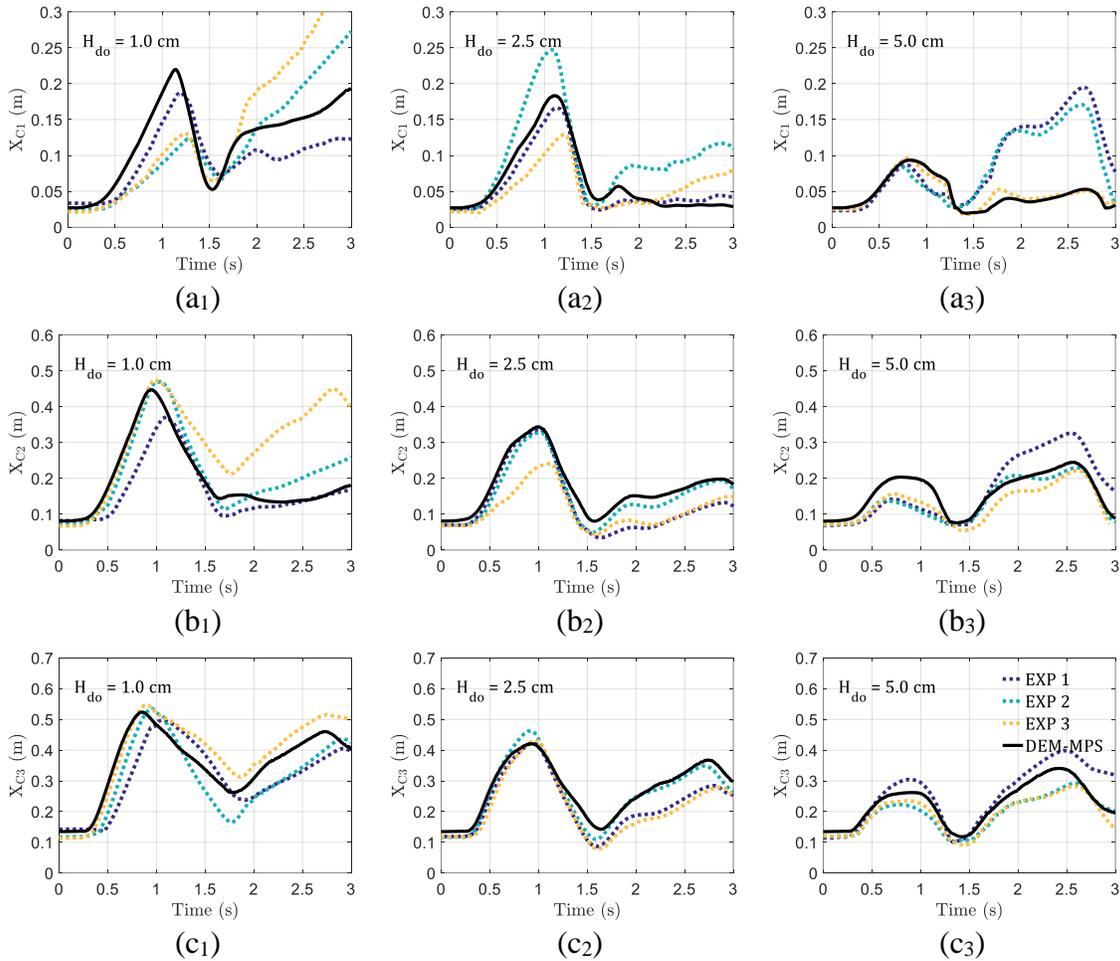

Fig. 24. Dam breaking with 9 blocks and wet beds: $H_{do}$ = 1, 2.5, and 5 cm. Experimental (dotted lines) and numerical (solid line) motions of the ($a_1$, $a_2$, $a_3$) block C1, ($b_1$, $b_2$, $b_3$) block C2 and ($c_1$, $c_2$, $c_3$) block C3 along the longitudinal direction.

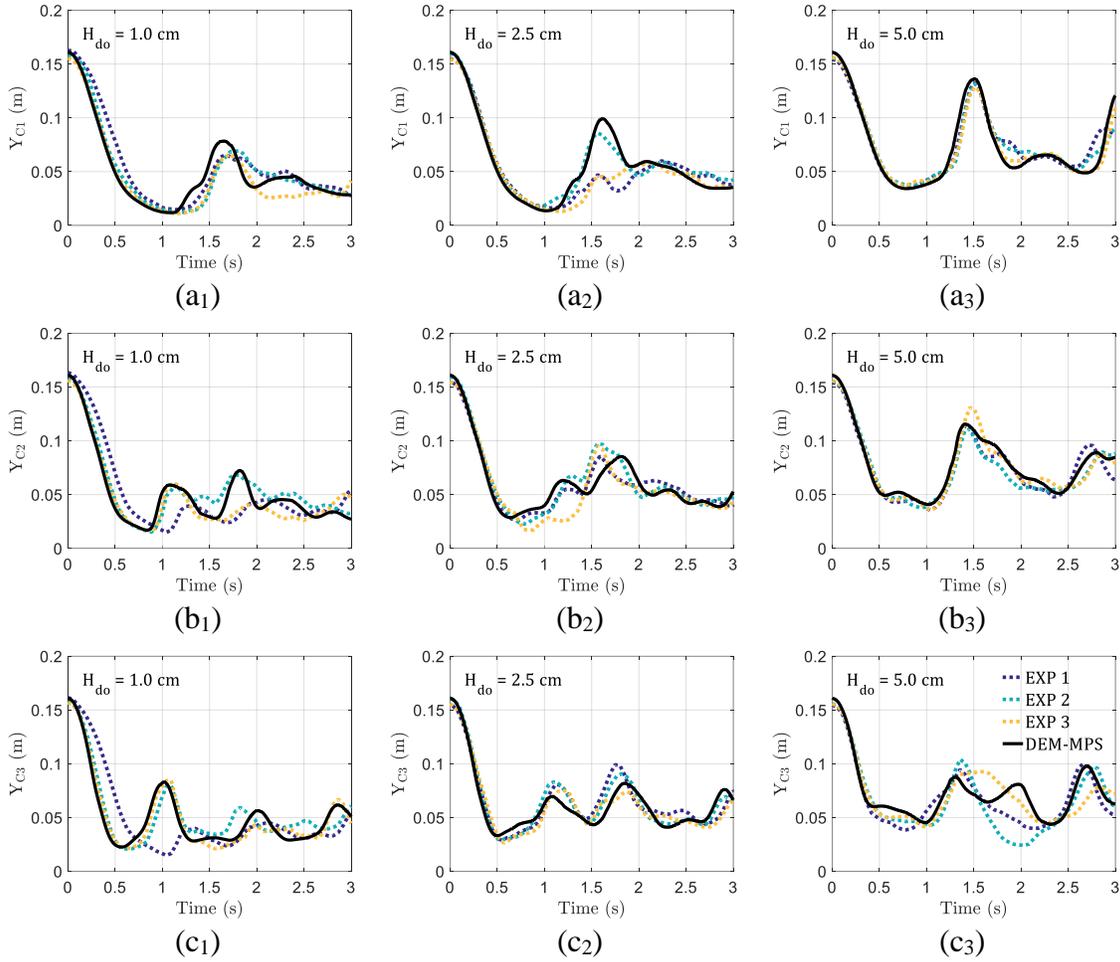

Fig. 25. Dam breaking with 9 blocks and wet beds: $H_{do}$ = 1, 2.5 and 5 cm. Experimental (dotted lines) and numerical (solid line) motions of the ($a_1$, $a_2$, $a_3$) block C1, ($b_1$, $b_2$, $b_3$) block C2 and ($c_1$, $c_2$, $c_3$) block C3 along the vertical direction.

## 6    Concluding remarks

A 3D fully Lagrangian numerical model, based on a hybrid DEM-MPS technique, for modeling the complex dynamics of ice-wave interaction was developed and evaluated in the present work. A set of benchmark dam-break experiments with floating block floes over dry and wet beds, which representing a simplified form of a jam release, were designed and conducted. The experiments provided quality data for the numerical validation and investigation of the dynamic behavior of ice floes and complex hydrodynamics.

Both the obtained experimental and numerical results show that the increase of initial downstream fluid level leads to a decrease of the wave front speed and the generation of the mushroom-like jet. These features are consistent with previous works (Stansby et al., 1998; Jánosi et al., 2004). In most cases, the maximum distances reached by the blocks during the dam-break phase are smaller when the downstream fluid layer is higher. However, this behavior is not observed for the experimental results of the A1 and B1 blocks (i.e. the further upstream blocks) in the dam-break phase with 4 blocks over the dry and wet bed of 1 cm fluid layer. The main reason for this discrepancy could be the difficulty in spacing equally and regularly the blocks apart.

The agreements between the experimental and numerical wave profiles as well as the positions of the blocks are quite satisfactory, thus, showing the effectiveness of the present approach to reproduce the main features of the fluid-solid interaction phenomenon. Nevertheless, the differences between computed and measured solid motions are more evident after the merging of the collapsed upward water flow and reflected wave. It is reasonable since the blocks are subjected to a very complex and highly non-linear free-surface flow during this stage.

In addition to the above, the experimental results provide comprehensive data for the validation and parametrization of theoretical/numerical models for modeling wave-ice floes interaction.

## 7    Limitations of the present model and future research

The results suggest that the present model can be extended to a parallelized solver able to handle simulations of real-scale ice dynamic problems with a large number of ice floes, such as ice-jam formation and breakup and ice floes interaction with hydraulic structures, and then supporting or even substituting laboratory experiments.

This study has been the first step toward developing a fully 3D Lagrangian model, for highly dynamic ice movements. The suggestion for future works will be to evaluate the model for test cases with real ice and extend the application of the model to study in-depth the physics of real-scale problems such as ice jam formation and breakup. Furthermore, a turbulent model is recommendable. Notwithstanding, one should keep in mind that the duration and effect of the shear forces near the tank bottom and lateral walls is attenuated when the upstream height $H_{up}$ and the water bed $H_{do}$ increase, as observed by Jánosi et al. (2004) from a series of experiments, and illustrated in Fig. 14 of the numerical study conducted by Khoshkonesh et al. (2019). Therefore, whereas the turbulence has an important effect in dry bed events, neglecting the turbulence does not imply considerable limitation of the proposed coupled model for some practical applications. Another limitation is the absence of a proper bed-to-ice friction model, which is very important near the wave front.

Since the dam breaking event is a gravity-driven phenomenon, the effects of the surface tension were neglected in numerical modeling. However, at the initial condition, the floating blocks, which have a large perimeter, were placed very close to each other, rendering the surface tension force strong enough to have relevant effects at the very beginning of the water column collapse, and possibly making it one of the causes of the chaotic nature of the experimental results.

The computed motions are well reproduced by the present model, but the computed forces are dependent on constants that require calibration, as the collision ratio and friction coefficient. In this sense, further effort is desirable to improve the model, such as the implementation of the solid-solid contact model based on the impulse method (Li et al., 2020), which requires less or no parameter tuning. Since shear forces would be very important in ice jam formation and release events, it is recommended to take into account fluid shear stress in close future works. Furthermore, a numerical model to simulate the internal ice stress and ice-breaking process, therefore suitable to reproduce drift ice dynamics, should also be considered in future studies.

**Acknowledgments**

Authors would like to acknowledge the support of the Coordenação de Aperfeiçoamento de Pessoal de Nível Superior - Brasil (CAPES) Finance Code 001 and the Natural Sciences and Engineering Research Council of Canada (NSERC) for providing scholarship support for the first and second authors, respectively. This research was also partially supported by Canada Research Chair Program. Authors would like to also thank Mr. Étienne Bélanger for his assistance in setting up the experiments. The first and fourth authors are also grateful to Petrobras for financial support on the development of the MPS/TPN-USP simulation system based on MPS method.

**Appendix A**

Table 5 shows the masses used for the calculation of shear forces and static friction coefficients between the blocks of polypropylene.

Table 5. Masses used to determine the coefficient of friction.

| Try | Mass (g) | Shear force (g) | Coefficient of Friction | Try | Mass (g) | Shear force (g) | Coefficient of Friction |
|---|---|---|---|---|---|---|---|
| 1 | 1702 | 675 | 0.397 | 10 | 2135 | 900 | 0.422 |
| 2 | 1702 | 700 | 0.411 | 11 | 2135 | 900 | 0.422 |
| 3 | 1702 | 675 | 0.397 | 12 | 2135 | 850 | 0.398 |
| 4 | 1702 | 650 | 0.382 | 13 | 2135 | 925 | 0.433 |
| 5 | 1702 | 750 | 0.441 | 14 | 2668 | 1050 | 0.394 |
| 6 | 2002.5 | 750 | 0.375 | 15 | 2668 | 1150 | 0.431 |

| 7 | 2002.5 | 825 | 0.412 | 16 | 2668 | 1225 | 0.459 |
| 8 | 2002.5 | 850 | 0.424 | 17 | 2668 | 1025 | 0.384 |
| 9 | 2002.5 | 900 | 0.449 | 18 | 2668 | 1100 | 0.412 |
| **Average** | | | | 0.413 | | | |
| **Median** | | | | 0.412 | | | |
| **Standard deviation** | | | | 0.023 | | | |

## Appendix B

In this appendix, a discussion of the effect of ice floes on the flow is outlined. For the sake of comparison, numerical simulations without blocks were also performed, and the computed wave profiles at selected instants are illustrated in Fig. 26 to Fig. 29. As can be noticed, the computed wave profiles without solids are slightly different from those of 4 blocks (Fig. 8, Fig. 9, Fig. 10 and Fig. 11) and 9 blocks (Fig. 18, Fig. 19, Fig. 20 and Fig. 21), and these small differences are more evident in the late stage of the hydrodynamics process, i.e., $t \geq 1.6$ s. A closer look at Fig. 8(d), Fig. 18(d), and Fig. 26(d) reveals that for the simulations with a dry bed, a slightly higher splash occurs for the case without blocks at $t = 1.6$ s. By comparing Fig. 9(f), Fig. 19(f), and Fig. 27(f), a very small difference for the run-up occurs between the simulations without and with 4 blocks for filling depth $H_{do} = 1$ cm at $t = 2.4$ s. Comparing the results for $H_{do} = 2.5$ cm, Fig. 10(d) and (f), Fig. 20(d) and (f), and Fig. 28(d) and (f) show that distinct splashes occur for the simulations without and with blocks at the instants $t = 1.6$ s and $t = 2.4$ s despite with relatively small difference. Finally, the comparison between Fig. 11(d), Fig. 21(d), and Fig. 29(d) illustrates that a higher splash is computed for the case without blocks and $H_{do} = 5$ cm at $t = 1.6$ s. In summary, these results show that the presence of the light-weight blocks is almost negligible in the initial gravity-dominant dam-break event. However, the presence of the blocks is not negligible as time goes by, and the intensity as well as the sequence of events, like the wave impact, run-up, and splashing, are higher in the cases without the blocks.

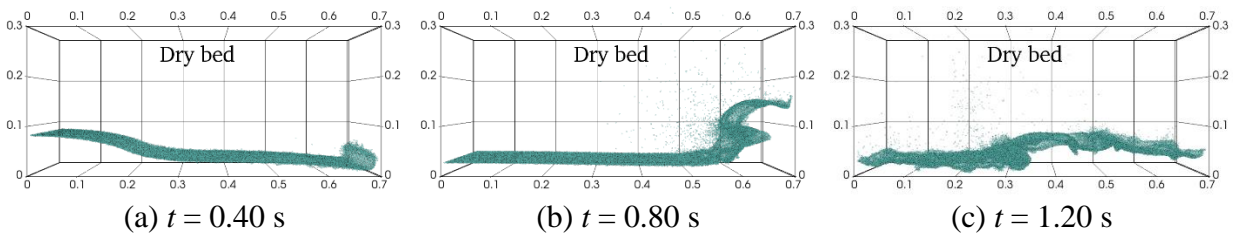

(a) $t = 0.40$ s    (b) $t = 0.80$ s    (c) $t = 1.20$ s

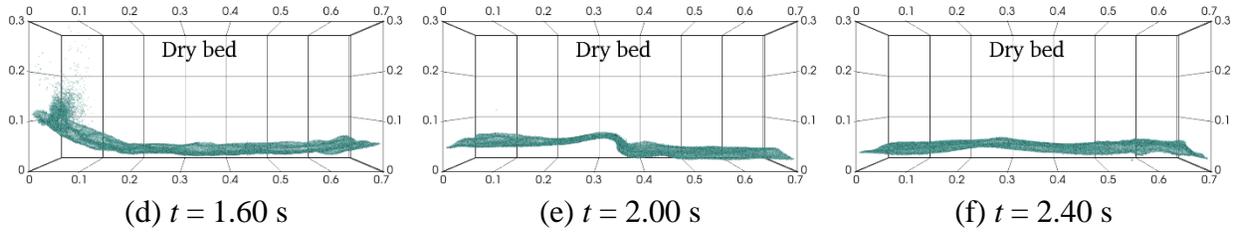

(d) $t = 1.60$ s        (e) $t = 2.00$ s        (f) $t = 2.40$ s

Fig. 26. Snapshots of the numerical dam breaking without blocks and dry bed at the instants $t = 0.40$, 0.80, 1.20, 1.60, 2.00, 2.40 s (front view).

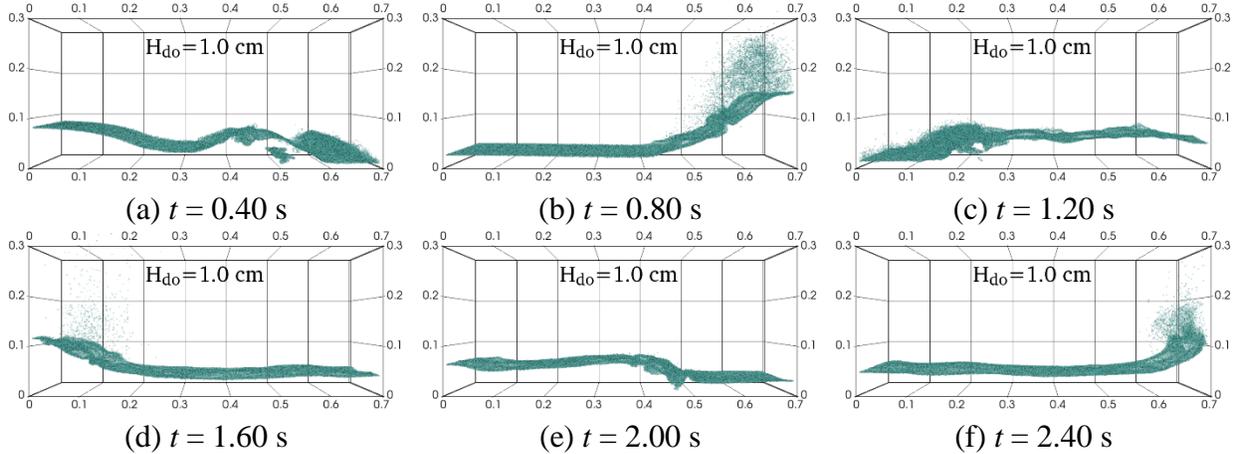

(a) $t = 0.40$ s        (b) $t = 0.80$ s        (c) $t = 1.20$ s

(d) $t = 1.60$ s        (e) $t = 2.00$ s        (f) $t = 2.40$ s

Fig. 27. Snapshots of the numerical dam breaking without blocks and wet bed $H_{do} = 1.0$ cm at the instants $t = 0.40$, 0.80, 1.20, 1.60, 2.00, 2.40 s (front view).

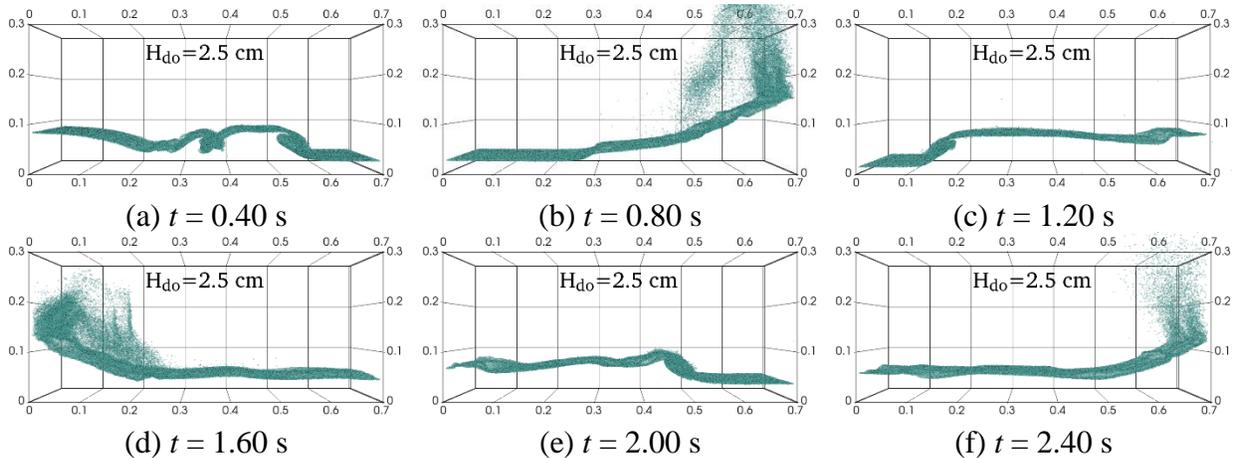

(a) $t = 0.40$ s        (b) $t = 0.80$ s        (c) $t = 1.20$ s

(d) $t = 1.60$ s        (e) $t = 2.00$ s        (f) $t = 2.40$ s

Fig. 28. Snapshots of the numerical dam breaking without blocks and wet bed $H_{do} = 2.5$ cm at the instants $t = 0.40$, 0.80, 1.20, 1.60, 2.00, 2.40 s (front view).

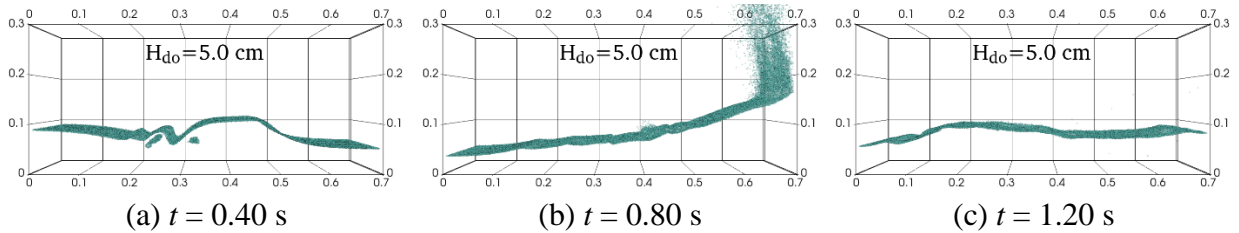

(a) $t = 0.40$ s        (b) $t = 0.80$ s        (c) $t = 1.20$ s

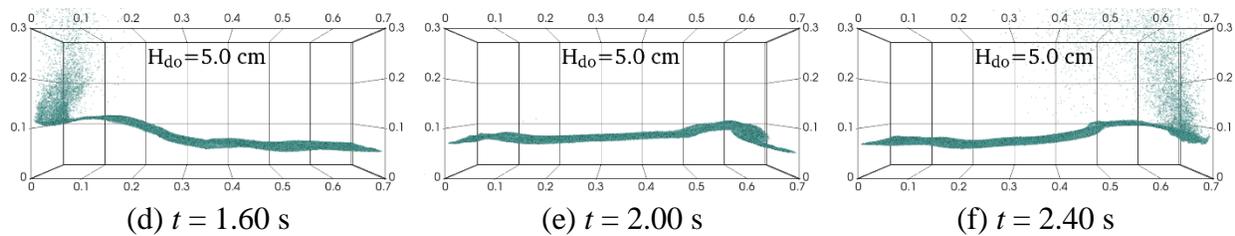

Fig. 29. Snapshots of the numerical dam breaking without blocks and wet bed $H_{do}$ = 5.0 cm at the instants $t$ = 0.40, 0.80, 1.20, 1.60, 2.00, 2.40 s (front view).